\documentclass[aps,pre,twocolumn,groupedaddress,floatfix,longbibliography,superscriptaddress]{revtex4-1}

\usepackage{amsmath}
\usepackage{amssymb}
\usepackage{amsfonts}
\usepackage{graphicx}
\usepackage{bbold}
\usepackage{bm}
\usepackage{bm}
\usepackage{cancel}
\usepackage{times,float}
\usepackage{graphicx}
\usepackage[usenames,dvipsnames,svgnames]{xcolor}
\usepackage{hyperref}
\hypersetup{colorlinks=true, linkcolor=Blue, citecolor=Blue,urlcolor=Blue}
\usepackage{multirow}
\usepackage{ulem}
\usepackage{color,epsfig}
\usepackage{bm}
\usepackage{slashed}
\usepackage{hyperref}
\usepackage{subfigure}
\usepackage{feynmf}
\usepackage{mathrsfs}
\usepackage{subcaption}

\newcommand{\bea}{\begin{eqnarray}}
\newcommand{\eea}{\end{eqnarray}}
\newcommand{\be}{\begin{equation}}
\newcommand{\ee}{\end{equation}}

\renewcommand\Re{\text{Re}}
\renewcommand\Im{\text{Im}}

\newcommand{\nn}{\nonumber}
\newcommand{\ii}{\mathrm{i}}

\newcommand{\qB}{|eB|}
\newcommand{\kt}{\mathbf{k}_\perp}
\newcommand{\pt}{\mathbf{p}_\perp}
\newcommand{\sign}{\text{sign}(eB)}

\newcommand{\kp}{k_\parallel}
\newcommand{\gt}{g_\perp}
\newcommand{\gp}{g_\parallel}

\newcommand{\gm}{\gamma^\mu}

\newcommand{\pp}{p_\parallel}

\newcommand{\Op}[1]{\mathcal{O}^{(#1)}}

\newcommand{\rp}{\rho_\parallel}
\newcommand{\rt}{\rho_\perp}
\newcommand{\B}{\mathcal{B}}

\begin{document}

\title{Fermion Self-Energy and Effective Mass in a Noisy Magnetic Background}
\author{Jorge David Casta\~no-Yepes}
\email{jcastano@uc.cl}
\affiliation{Facultad de F\'isica, Pontificia Universidad Cat\'olica de Chile, Vicu\~{n}a Mackenna 4860, Santiago, Chile}
\author{Enrique Mu\~noz}
\email{ejmunozt@uc.cl}
\affiliation{Facultad de F\'isica, Pontificia Universidad Cat\'olica de Chile, Vicu\~{n}a Mackenna 4860, Santiago, Chile}
\affiliation{Center for Nanotechnology and Advanced Materials CIEN-UC, Avenida Vicuña Mackenna 4860, Santiago, Chile}

\begin{abstract}
In this article, we consider the propagation of QED fermions in the presence of a classical background magnetic field with white-noise stochastic fluctuations. The effects of the magnetic field fluctuations are incorporated into the fermion and photon propagators in a quasi-particle picture, which we developed in previous works using the {\it replica trick}. By considering the strong-field limit, here we explicitly calculate the fermion self-energy involving radiative contributions at first-order in $\alpha_\text{em}$, in order to obtain the noise-averaged mass of the fermion propagating in the fluctuating magnetized medium. Our analytical results reveal a leading double-logarithmic contribution $\sim \left[\ln \left( |e B|/m^2  \right)\right]^2$ to the mass, with an imaginary part representing a spectral broadening proportional to the magnetic noise auto-correlation $\Delta$.
While a uniform magnetic field already breaks Lorentz invariance, inducing the usual separation into two orthogonal subspaces (perpendicular and parallel with respect to the field), 
the presence of magnetic noise further breaks the remaining symmetry, thus leading to distinct spectral widths associated with fermion and anti-fermion, and their spin projection in the quasi-particle picture.
\end{abstract}

\maketitle 
\section{Introduction}
The presence of strong, transient and inhomogeneous magnetic fields and their effects over elementary particles is a subject of great interest in several physical scenarios, such as ultrarelativistic heavy-ion collisions\cite{Alam_2021,ayala2022anisotropic,Inghirami_2020,Ayala:2019jey,Ayala:2017vex} and the corresponding genesis of the quark-gluon plasma~\cite{Busza_2018,Hattori_2016,Hattori_2018,Buballa_2005}. In such a magnetized medium, fermions (charged leptons and quarks) as well as neutral gauge fields (photons and gluons) develop non-trivial responses due to vacuum polarization effects.  
The case of a classical, static and homogeneous background magnetic field, since the seminal work of Schwinger~\cite{Schwinger_1951}, has been discussed extensively in the literature~\cite{Dittrich_Reuter,Dittrich_Gies}. This idealized scenario has been studied in the context of the gluon polarization tensor~\cite{Hattori_2013,Hattori_2016,Ayala_Pol_020,Hattori_2018}, where the presence of the magnetic field breaks the Lorentz invariance, thus generating three tensor components with their corresponding refraction indexes, the so-called vacuum birefringence phenomena~\cite{Hattori_2013,Ayala_Pol_020}. Similarly, the propagation of fermions in such a uniform magnetized background~\cite{Tsai_PhysRevD.10.1342,Ayala_mass_021}, as expressed by the self-energy, leads to the definition of a magnetic mass with a leading double-logarithmic dependence $\sim \left[\ln \left( |e B|/m^2  \right)\right]^2$, where $|eB|$ and $m$ are the background field and bare fermion mass, respectively~\cite{Machet_10.1142/S0217751X16500718,Jancovici_PhysRev.187.2275,Ayala_mass_021,Dittrich_Reuter}. In addition, an spectral broadening arising from imaginary parts in the fermion self-energy was recently predicted due to Landau level mixing~\cite{Ayala_mass_021}. 
Since spatio-temporal fluctuations in the background magnetic field should indeed exist in the physical scenarios of interest~\cite{Inghirami_2020}, we recently developed a theoretical formalism to include random fluctuations with respect to a uniform and constant magnetic field background, as a next step towards a more realistic approximation~\cite{Castano_Munoz_PhysRevD.107.096014}. For the gauge fields $A^{\mu}(x)$, we distinguish three physically different contributions
\begin{eqnarray}
A^{\mu}(x) \rightarrow A^{\mu}(x) + A^{\mu}_\text{BG}(x) + \delta A^{\mu}_\text{BG}(\mathbf{x}),
\label{eq_Atot}
\end{eqnarray}
where $A^{\mu}(x)$ is the dynamical photonic quantum field, while BG represents the classical ``background''. Moreover, for this BG contribution, we consider the effect of white noise spatial fluctuations $\delta A^{\mu}_\text{BG}(\mathbf{x})$ with respect to the mean value $A_\text{BG}^{\mu}(x)$, satisfying the statistical properties
\begin{eqnarray}
\langle \delta A^{j}_\text{BG}(\mathbf{x}) \delta A^{k}_\text{BG}(\mathbf{x}')\rangle &=&  \Delta\delta_{j,k}\delta^{3}(\mathbf{x}-\mathbf{x}'),\nonumber\\
\langle \delta A^{\mu}_\text{BG}(\mathbf{x})\rangle &=& 0,
\label{eq_Acorr}
\end{eqnarray}
such that the corresponding classical magnetic field background is $\mathbf{B} + \delta\mathbf{B}(x) = \nabla\times\left( \mathbf{A}_\text{BG}(x) + \delta \mathbf{A}_\text{BG}(x) \right)$, with a uniform mean value $\mathbf{B}$.

In our formalism, a statistical average is performed over a functional distribution of random fluctuations $\delta \mathbf{A}_\text{BG}(x)$ of the classical gauge fields. This procedure is mathematically implemented by means of the replica trick~\cite{Parisi_Mezard_1991,Kardar_Parisi_1986}, and as a result it provides explicit analytical expressions for the fermion and gauge field propagators, respectively, as a perturbative expansion in terms of the magnetic noise autocorrelation $\Delta \ge 0$. We have shown that the effects of magnetic noise over the fermion propagator is equivalent to a dispersive media, with an effective refraction index that modifies the group velocity of the otherwise free particles, but not their mass~\cite{Castano_Munoz_PhysRevD.107.096014}. In contrast, we have recently shown~\cite{Castano_Munoz_PhysRevD.109.056007} that as a consequence of the magnetic fluctuations combined with charged vacuum polarization in QED, photons develop anisotropic magnetic masses $M_{\perp}$ and $M_{\parallel}$ which are proportional to $\Delta$~\cite{Castano_Munoz_PhysRevD.109.056007}. Therefore, a related question remains opened: What are the effects of the background magnetic noise over the fermion self-energy and its corresponding effective mass, when radiative effects are taken into account? In this article, we shall address this question from a perturbation theory perspective within the framework of QED, by applying our previous results for the noise-averaged fermion and photon propagators, respectively, in the very strong magnetic field limit $|e B|\gg m^2$ which is relevant for heavy-ion collisions.

\section{The QED fermion self-energy at 1-loop}
In the configuration space, for a QED fermion with charge $-e$ and mass $m$ propagating through a magnetized medium, its
self-energy due to radiative effects at 1-loop (as depicted in Fig.~\ref{fig:selfenergy}) can be expressed as follows:
\bea
-\ii\Sigma(x,x')=(-\ii e)^2\gamma^\mu \ii S(x,x')\gamma^\mu D_{\mu\nu}(x-x').
\eea
Here, the fermion propagator is given by 
    \bea
    \ii S(x,x')=\Phi(x,x')\int\frac{d^4k}{(2\pi)^4}e^{-\ii k\cdot(x-x')}\ii S(k),
    \eea
while 
    \bea
    D_{\mu\nu}(x-x')=\int\frac{d^4q}{(2\pi)^4}e^{-\ii q\cdot(x-x')}D_{\mu\nu}(q)
    \eea
represents the photon propagator.

As discussed in the literature~\cite{Schwinger_1951,Dittrich_Reuter}, the presence of the magnetic field breaks the translational invariance of the propagators, but gauge-covariance is granted by the Schwinger phase factor $\Phi(x,x')$, which takes the form:
    \bea
    \Phi(x,x')=\exp\Bigg\{\ii e\int_x^{x'}d\xi^\mu\left[A_\mu+\frac{1}{2}F_{\mu\nu}(\xi-x')^\nu\right]\Bigg\}.
    \eea

For a magnetic field $\mathbf{B}$ oriented along the $\hat{z}$-direction, in the symmetric gauge
    \bea
    A_\mu=\frac{B}{2}(0,-x_2,x_1,0),
    \eea
we obtain the phase explicitly as
    \bea
    \Phi(x,x')=\exp\left(\frac{\ii eB}{2}\epsilon_{ij}x_ix_j'\right),
    \eea
    where $\epsilon_{ij}$ is the two dimensional Levy-Civita tensor. 

Similarly, and in consistency with the Dyson equation for the fermion propagator in configuration space, the fermion self-energy also involves the Schwinger phase 
    \bea
    -\ii\Sigma(x,x')=\Phi(x,x')\int\frac{d^4p}{(2\pi)^4}e^{-\ii p\cdot(x-x')}\left[-\ii\Sigma(p)\right],
    \eea
where the translational-invariant factor is given by the expression
    \bea
    -\ii\Sigma(p)\equiv(-\ii e)^2\int\frac{d^4k}{(2\pi)^4}\gamma^\mu\ii S(k)\gamma^\nu D_{\mu\nu}(p-k).
    \eea

The presence of a constant magnetic field background breaks the Lorentz invariance. Therefore, the phase space is splitted into two subspaces according to the parallel and perpendicular directions with respect to the background field.  The metric tensor is thus splitted accordingly $g^{\mu\nu}=\gp^{\mu\nu}+\gt^{\mu\nu}$,
where
\bea
\gp^{\mu\nu}&=&\text{diag}(1,0,0,-1)\nn\\
\gt^{\mu\nu}&=&\text{diag}(0,-1,-1,0).
\eea
The latter implies that for any four-vector
\bea
    p^\mu&=&\pp^\mu+p_\perp^\mu,
\eea
the inner product also splits accordingly $p^2=\pp^2-\pt^2$,
with
\bea
\pp^2&=& \gp^{\mu\nu}p_{\mu}p_{\nu} =  p_0^2-p_3^2\nn\\
p_{\perp}^2 &=& \gt^{\mu\nu}p_{\mu}p_{\nu} = -  \pt^2= -( p_1^2+p_2^2 ).
\eea

\subsection{The noise-averaged propagators}
In two of our recent articles, we showed that the fermion propagator in a media with fluctuations in a external and very intense magnetic field can be expressed as~\cite{Castano_Munoz_PhysRevD.107.096014,Castano_Munoz_PhysRevD.109.056007}:
 \bea
\ii S(p)&=&\ii S_0(p)+ \Delta\cdot\ii  S_1(p)+O(\Delta^2),
\eea
where the noiseless part of the fermion propagator is provided by its expression in the so-called Lowest Landau Level (LLL)\cite{Fukushima_PhysRevD.83.111501,Bandyopadhyay_PhysRevD.94.114034}:
\bea
\ii S_0(p)=2\ii\frac{e^{-\pt^2/\qB}}{\pp^2-m^2+\ii\epsilon}(\slashed{p}_{\parallel}+m)\mathcal{O}^{(\uparrow)},
\label{eq:S0def}
\eea
and the noise-averaged part is given by~\cite{Castano_Munoz_PhysRevD.107.096014,Castano_Munoz_PhysRevD.109.056007}
 \bea
    \ii S_1(p)&\equiv&\ii\left(\frac{\qB}{2\pi}\right)\Big[\Theta_1(p)(\slashed{p}_\parallel+m)\mathcal{O}^{(\uparrow)}\nn\\
&-&\Theta_2(p)\gamma^3\mathcal{O}^{(\uparrow)}+\Theta_3(p)\sign\ii\gamma^1\gamma^2(\slashed{p}_\parallel+m)\Big],\nn\\
\label{eq:SDdef}
    \eea
where
\begin{subequations}
\bea
\mathcal{O}^{(\uparrow,\downarrow)}=\frac{1}{2}\left[1\mp\text{sign}(eB)\ii\gamma^1\gamma^2\right],
\eea
    \bea
\Theta_1(p)&\equiv&\frac{3(\pp^2+m^2)e^{-2\pt^2/\qB}}{(\pp^2-m^2)^2\sqrt{p_0^2-m^2}} ,
\label{Theta1}
\eea
\bea
\Theta_2(p)&\equiv&\frac{3p_3e^{-2\pt^2/\qB}}{(\pp^2-m^2)\sqrt{p_0^2-m^2}},
\label{Theta2}
\eea
\bea
\Theta_3(p)&\equiv&\frac{e^{-2\pt^2/\qB}}{(\pp^2-m^2)\sqrt{p_0^2-m^2}}.
\label{Theta3}
\eea

\end{subequations}

On the other hand, in our recent work~\cite{Castano_Munoz_PhysRevD.109.056007} we calculated the photon propagator in the same strong magnetic field regime, such that its average over magnetic noise takes the form:
\bea
D^{\mu\nu}(q)&=&\frac{-\ii g_\parallel^{\mu\nu}}{q^2+\ii M_\parallel^2+\ii\epsilon}+\frac{-\ii g_\perp^{\mu\nu}}{q^2-\ii M_\perp^2+\ii\epsilon}\nn\\
&-&\frac{2 M_\perp^2~\delta_3^\mu\delta_3^\nu}{\left(q^2+\ii M_\parallel^2+\ii\epsilon\right)\left(q^2+\ii (M_\parallel^2-3M_\perp^2)+\ii\epsilon\right)},\nn\\
\label{eq_photpropfull}
\eea
thus revealing that photons in such dispersive media may acquire magnetic masses proportional to the noise $\Delta$, i.e. $M_\parallel$, and $~M_\perp$ given by~\cite{Castano_Munoz_PhysRevD.109.056007}:
\bea
M_\parallel^2\equiv\frac{59\alpha_\text{em}|eB|^2\Delta}{96\pi m};~M_\perp^2\equiv\frac{\alpha_\text{em}|eB|^2\Delta}{3\pi m},
\eea
where $\alpha_\text{em}\equiv\frac{e^2}{4\pi}$ is the QED fine structure constant. For practical purposes, we shall expand the propagator in Eq.~\eqref{eq_photpropfull} up to first-order in $\Delta$
\bea
D^{\mu\nu}(q)&=&D^{\mu\nu}_0(q)+\Delta\cdot D^{\mu\nu}_1(q)+O(\Delta^2),
\label{eq_photpropser}
\eea
with
\bea
D^{\mu\nu}_0(q)=\frac{-\ii g^{\mu\nu}}{q^2+\ii\epsilon},
\label{eq:D0def}
\eea
the free photon propagator in the Feynman gauge, 
and
\bea
D^{\mu\nu}_1(q)=-\frac{\alpha_\text{em}\qB^2}{96\pi m(q^2+\ii\epsilon)^2}\left(59 g_\parallel^{\mu\nu}-32g_\perp^{\mu\nu}+64\delta_3^\mu\delta_3^\nu\right).\nn\\
\label{eq:DDdef}
\eea
the first order contribution due to the magnetic background noise.

\subsection{The noise-averaged fermion self-energy}
\begin{figure}[b!]
    \centering
    \includegraphics[width=0.48\textwidth]{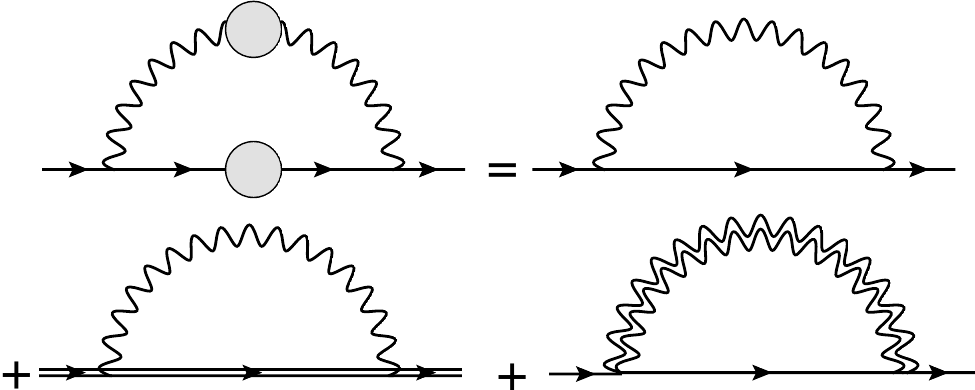}
    \caption{Feynman diagrams depicting the contributions to fermion self-energy up to order $O(\Delta^2)$. The single lines represent fermions and photons in a constant and intense magnetic field, while double lines depict fermions and photons in a fluctuating background magnetic field. }
    \label{fig:selfenergy}
\end{figure}

From a perturbation theory analysis, and considering the system subjected to a uniform magnetic filed background as a reference, the self-energy averaged over magnetic noise, up to first-order in the noise auto-correlation $\Delta$, is expressed by the following terms at first order in $\alpha_\text{em}$ and $\Delta$:
\bea
\Sigma(p,B,\Delta)&=&\Sigma_{0}^r(p,B)+\Sigma_{\Delta}^{r}(p,B)+O(\Delta^2).
\eea

These contributions are represented by the Feynman diagrams depicted in Fig.~\ref{fig:selfenergy}, and defined by the following algebraic expressions, plus the corresponding counterterms
\begin{subequations}
    \bea
    -\ii\Sigma_{0}(p,B)\equiv(-\ii e)^2\int\frac{d^4k}{(2\pi)^4}\gamma^\mu\ii S_0(k)\gamma^\nu D_{\mu\nu}^0(p-k),\nn\\
    \label{eq:Sigma00def}
    \eea
    \bea
    -\ii\Sigma_{\Delta}(p,B)\equiv(-\ii e)^2 \Delta \int\frac{d^4k}{(2\pi)^4}\gamma^\mu\ii S_1(k)\gamma^\nu D_{\mu\nu}^0(p-k),\nn\\
    \label{eq:SigmaD0def}
    \eea
    \label{eq:Sigmaij}
\end{subequations}
    %
    
Even though there are three Feynman diagrams depicted in Fig.~\ref{fig:selfenergy}, the third one is $O(\alpha_\text{em}^2)$ and hence is not included in our approximation for overall consistency in perturbation theory.

The counter-terms must be determined in order to impose the physically appropriate renormalization conditions. We remark that in the noiseless limit $\Delta =0$, the full Lorentz invariance of the propagator is already broken by the presence of the uniform background magnetic field, that here we assume is very intense ($|eB|\gg m^2$). In such case, the {\it{magnetic mass operator}}, as already discussed in our previous work~\cite{Ayala_mass_021}, is given by the expression
\bea
\hat{M}_B(\Delta = 0) = m + \left.\Sigma_{0}(p,B)\right|_{\substack{\slashed{p}_{\parallel}=m,\\\slashed{p}_{\perp}=0}}.
\label{eq:mB0}
\eea

On the other hand, we remark that (as seen already in the structure of the propagators Eq.~\eqref{eq:SDdef} and Eq.~\eqref{eq:DDdef}), the presence of magnetic noise fully breaks the Lorentz symmetry, such that when $\Delta > 0$ we are restricted to define the physical mass from the condition $\slashed{p}_{\parallel} = \gamma^0 m$, $\mathbf{p} = 0$. Therefore, in order for the limit $\Delta\rightarrow 0^{+}$ to correctly match the noiseless result in Eq.~\eqref{eq:mB0}, we consider the following renormalization conditions
\begin{subequations}
\bea
\lim_{\Delta\rightarrow 0^+}\left.\Sigma(p,B,\Delta)\right|_{\slashed{p}_{\parallel} = m\gamma^0,\mathbf{p}=0} &=& \left.\Sigma_{0}^{r}(p,B)\right|_{\slashed{p}_{\parallel}=m\gamma^0,\mathbf{p}=0}\nn\\
&=& \hat{M}_B(\Delta=0) - m\nn\\
&=& \left.\Sigma_{0}(p,B)\right|_{\slashed{p}_{\parallel}=m,\slashed{p}_{\perp}=0}\nn\\
\eea
and
\bea
&\lim_{\Delta\rightarrow 0^+}\left.\frac{\partial}{\partial\slashed{p}_{\parallel}}\Sigma(p,B,\Delta)\right|_{\substack{\slashed{p}_{\parallel} = m\gamma^0,\\\mathbf{p}=0}}\nn\\
&= \left.\frac{\partial}{\partial\slashed{p}_{\parallel}}\Sigma_{0}^{r}(p,B)\right|_{\substack{\slashed{p}_{\parallel}=m\gamma^0,\\\mathbf{p}=0}} = \left.\frac{\partial}{\partial\slashed{p}_{\parallel}}\Sigma_{0}(p,B)\right|_{\substack{\slashed{p}_{\parallel}=m,\\ \mathbf{p}_{\perp}=0}}
\eea
\end{subequations}
that ensure that the pole and the residue of the propagator correspond to the physical fermion mass in the limit of zero magnetic noise, but finite background field.  

These conditions allow us to determine the corresponding counterterms $\delta_Z$ and $\delta_m$ in the renormalized expression
\bea
\Sigma_{0}^r(p,B) = \Sigma_{0}(p,B) + \delta_Z (\slashed{p}_{\parallel}-m\gamma^0) - \delta_m, 
\eea
such that we have
\bea
\delta_m &=& \Sigma_{0}(\slashed{p}_{\parallel}=m\gamma^0,\mathbf{p}=0,B) - \Sigma_{0}(\slashed{p}_{\parallel}=m,\mathbf{p}_{\perp}=0,B),\nn\\
\delta_{Z} &=& \left.\frac{\partial}{\partial\slashed{p}_{\parallel}}\Sigma_{0}(p,B)\right|_{\substack{\slashed{p}_{\parallel}=m,\\\mathbf{p}_{\perp}=0}} - \left.\frac{\partial}{\partial\slashed{p}_{\parallel}}\Sigma_{0}(p,B)\right|_{\substack{\slashed{p}_{\parallel}=m\gamma^0,\\\mathbf{p}=0}}
\eea

The expressions for $-\ii\Sigma_{i}(p,B)$ for $i=0, \Delta$ are computed in Appendixes~\ref{Ap:Sigma00} and~\ref{Ap:SigmaDelta}, in order to determine the magnetic mass operator $\hat{M}_B(\Delta)$ according to the definition and renormalization prescriptions described above, as will be shown in the next section.

\section{The noise-averaged magnetic mass operator at first order in $\alpha_\text{em}$ and $\Delta$}

The noise-averaged magnetic mass operator of the fermion is obtained from the corresponding expression for the self-energy at $\slashed{p}_{\parallel} = \gamma^0 m$, $\mathbf{p} = 0$, as discussed in the previous section
\bea
\hat{M}_B(\Delta) - m
&=& \left.\Sigma(p,B,\Delta)\right|_{\slashed{p}_{\parallel}=m\gamma^0,\mathbf{p}=0}\\
&=& \left.\Sigma_{0}^{r}(p,B)\right|_{\substack{\slashed{p}_{\parallel}=m\gamma^0,\\\mathbf{p}=0}}+ \left.\Sigma_{\Delta}^r(p,B)\right|_{\substack{\slashed{p}_{\parallel}=m\gamma^0,\\\mathbf{p}=0}}\nn\\
&=& \hat{M}_B(\Delta=0) - m+\left.\Sigma_{\Delta}^r(p,B)\right|_{\substack{\slashed{p}_{\parallel}=m\gamma^0,\\\mathbf{p}=0}}\nn
\eea
where the magnetic mass operator in the noiseless limit $\Delta = 0$, as shown in detail in Appendix~\ref{Ap:Sigma00}, is given by
\bea
\hat{M}_B(\Delta = 0) - m &=& \left.\Sigma_{0}(p,B)\right|_{\slashed{p}_{\parallel}=m,\mathbf{p}_{\perp}=0}\nn\\
&=& \Op{\uparrow}M_{B0}^{(\uparrow)} + \Op{\downarrow}M_{B0}^{(\downarrow)},
\eea
where we defined the fermion magnetic mass eigenvalues for the $\uparrow$ and $\downarrow$ spin projections
by
\bea
M_{B0}^{(\uparrow)} &=& \frac{\alpha_\text{em}m}{\pi}\left[   \ln^2\B - \left( \gamma_e + \ii\frac{\pi}{2}\right)\ln\B+ \frac{\pi^2}{3}  \right]
+ O(\B^{-1})\nn\\
M_{B0}^{(\downarrow)} &=& \frac{\alpha_\text{em}m}{\pi}\left[   \ln^2\B - \left(1 + \gamma_e + \ii\frac{\pi}{2}\right)\ln\B\right.\nn\\
&&\left.-\left( 2 - \gamma_e - \frac{\pi^2}{3} - \ii\frac{\pi}{2}  \right)  \right]
+ O(\B^{-1}),
\label{eq:mB02}
\eea

The contribution arising from the self-energy terms proportional to the magnetic noise autocorrelation $\Delta$, as shown in detail in Appendix~\ref{Ap:SigmaDelta}, are defined in terms of the two projectors, 
\bea
\mathcal{P}^{(\pm)}\equiv\frac{1}{2}\left(\mathbb{1}\pm\gamma^0\right)
\eea
onto the fermion ($+$) and anti-fermion ($-$) subspaces, respectively, in the rest-frame $\mathbf{p} = 0$.

Therefore, we can split the noise contribution to the self-energy into four subspaces, namely
\bea
\lim_{\substack{p_0\to m\\\mathbf{p}\to 0}}\left[-\ii\Sigma_{\Delta}(p)\right]_r=\sum_{\sigma=\uparrow,\downarrow}\sum_{\lambda=\pm1}\left[-\ii\widetilde{\Sigma}_{\Delta}^{(\sigma,\lambda)}\Op{\sigma}\mathcal{P}^{(\lambda)}\right],\nn\\
\eea
where we defined the coefficients:
\begin{subequations}
   \bea
\widetilde{\Sigma}_{\Delta}^{(\downarrow,\pm)}\equiv\ii\frac{\sqrt{2}}{\pi^{3/2}}\left(3\ln(2)\pm8\right)\alpha_\text{em}\Delta\sqrt{\qB} m+O(m^2),\nn\\
\eea 
   \bea
\widetilde{\Sigma}_{\Delta}^{(\uparrow,\pm)}\equiv\ii\frac{\sqrt{2}}{\pi^{3/2}}\left(3\ln(2)\mp2\right)\alpha_\text{em}\Delta\sqrt{\qB} m+O(m^2),\nn\\
\eea 
\end{subequations}
which are clearly purely imaginary. 

\begin{figure}
    \centering
    \includegraphics[scale=0.75]{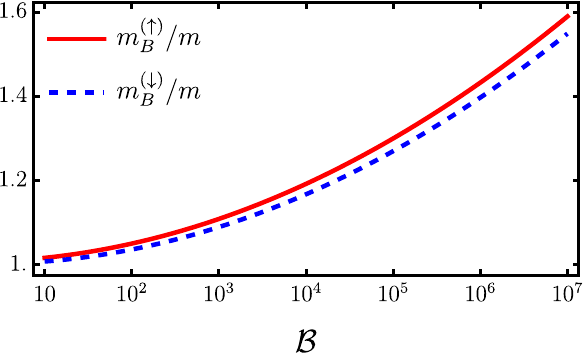}
    \caption{The magnetic mass of the fermion, calculated from Eqs.~\eqref{eq:mBreal}, is shown for the two spin projections as a function of the average background magnetic field (in dimensionless units) $\B = |eB|/m^2$.}
    \label{fig:Re_m}
\end{figure}

In summary, our results indicate that the fermion magnetic mass operator, in the presence of noise $\Delta>0$, possesses four different eigenvalues depending on the spin $\sigma = \uparrow,\downarrow$ and $\lambda=\pm$ projections, as follows
\bea
M_B^{(\sigma,\lambda)}(\Delta) = m + M_{B0}^{(\sigma)} + \widetilde{\Sigma}_{\Delta}^{(\sigma,\lambda)}.
\eea

We notice that these eigenvalues are complex, such that the real parts strictly correspond to the fermion magnetic mass, i.e. $m_B^{(\sigma)} = \Re\,M_B^{(\sigma,\lambda)}(\Delta)$, which turns out to be noise-independent
\bea
m_B^{(\uparrow)} &=& m + \frac{\alpha_\text{em}m}{\pi}\left[   \ln^2\B -  \gamma_e \ln\B+ \frac{\pi^2}{3}  \right] + O(\B^{-1})\nn\\
m_B^{(\downarrow)} &=& m + \frac{\alpha_\text{em}m}{\pi}\left[   \ln^2\B - (1 +  \gamma_e)\ln\B\right.\nn\\
&&\left.+ \frac{\pi^2}{3} + \gamma_e - 2  \right] + O(\B^{-1}),
\label{eq:mBreal}
\eea
which are depicted in Fig.~\ref{fig:Re_m}. We notice that the magnetic mass is different for each spin projection, as expected from the Zeeman interaction splitting. This effect becomes stronger in very intense magnetic fields, and may be of interest in different physical scenarios.

On the other hand, the imaginary parts represent  a Breit-Wigner resonance $\Gamma^{(\sigma,\lambda)}(\Delta) = -2\Im\, M_{B}^{(\sigma,\lambda)}(\Delta)$ due to the combination of the field and the magnetic noise, given by
{\small
\begin{subequations}
  \bea
\Gamma^{(\uparrow,\pm)}(\Delta) = \alpha_\text{em}m\left(\ln\B - \frac{2\sqrt{2\B}m\Delta}{\pi^{3/2}}\left(3\ln(2)\mp 2\right) \right),\nn\\
\eea
\bea
\Gamma^{(\downarrow,\pm)}(\Delta) = \alpha_\text{em}m\Bigg(\ln\B - 1 -   \frac{2\sqrt{2\B}m\Delta}{\pi^{3/2}}\left(3\ln(2)\pm 8\right) \Bigg).\nn\\
\eea  
\label{eq:mBupdown}
\end{subequations}
}

\begin{figure}[h!]
    \centering
    \includegraphics[scale=0.75]{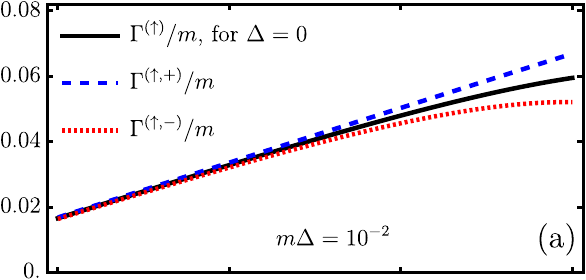}\nn\\
    ~\includegraphics[scale=0.755]{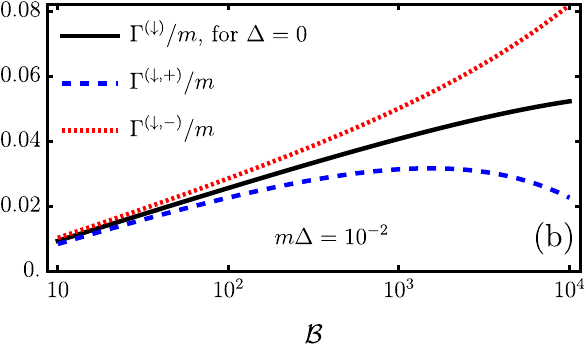}\\
    \vspace{0.5cm}
    \includegraphics[scale=0.75]{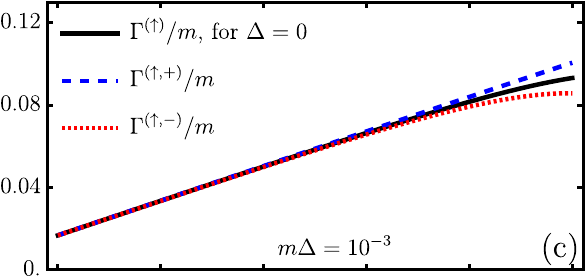}\nn\\
    ~\includegraphics[scale=0.755]{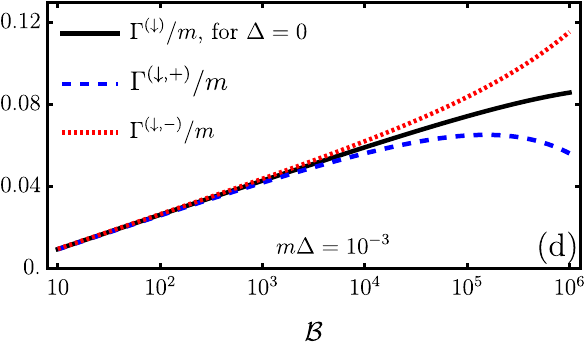}
    \caption{Breit-Wigner resonance $\Gamma^{(\sigma,\lambda)}$ computed from Eqs.~\eqref{eq:mBupdown}, as a function of the average background field $\mathcal{B} = |eB|/m^2$. The eigenvalues corresponding to projections onto the four independent sub-spaces are shown for two different values of the magnetic noise auto-correlation $m\Delta$. }
    \label{fig:ImUp}
\end{figure}

Figure~\ref{fig:ImUp} shows the behaviour of $\Gamma^{(\sigma,\lambda)}(\Delta)$, computed from Eqs.~\eqref{eq:mBupdown}, as a function of the average magnetic field (in dimensionless units) $\mathcal{B} = |eB|/m^2$, for the four eigenvalues corresponding to each projection $(\uparrow\downarrow,\pm)$, respectively. As can be seen in the figures, deviations from the noiseless limit $\Delta=0$ (solid line)  become appreciable after some critical value $\B > \mathcal{B}_c$ that depends on the magnitude of $m\Delta$ via the product $m\Delta\,\sqrt{\B}$. In cases where the spin projection is parallel ($\uparrow$) to the direction of the background magnetic field, the imaginary part of the magnetic mass in the sub-space given by the projection $\mathcal{P}^{(+)}$  decreases as compared to the corresponding one for the projection $\mathcal{P}^{(-)}$, and also with respect to the noiseless case $\Delta = 0$. The opposite occurs when the spin projection is anti-parallel to the direction of the background magnetic field. This implies that in the quasi-particle picture, the charge conjugation combined with the breaking of Lorentz symmetry provided by the magnetic noise results in different spectral widths for the various modes.

Note that in the heavy-ion collisions scenario, the magnetic background is about the pion-mass squared, so that for electrons we would have $\mathcal{B}\sim8\times10^4$, while for light quarks $\mathcal{B}\sim 8\times 10^3$. Therefore, our approximation based on the LLL expression for the fermion propagator valid for $\B \gg 1$ is well justified. Hence, for some ranges of the noise $m\Delta$, the effects displayed in Fig.~\ref{fig:ImUp} might be detected in actual experiments.

\begin{figure}
    \centering
    \includegraphics[scale=0.75]{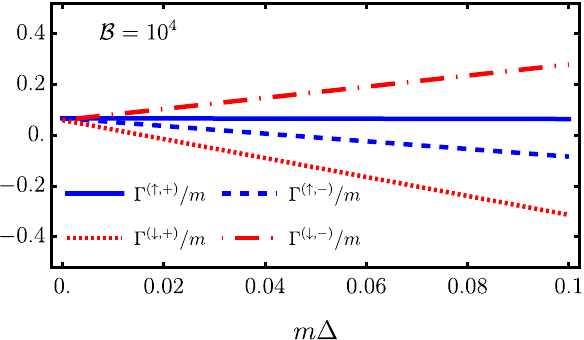}
    \caption{The Breit-Wigner resonance $\Gamma^{(\sigma,\lambda)}(\Delta)$, computed from Eqs.~\eqref{eq:mBupdown}, as a function of the noise auto-correlation parameter $m\Delta$, for a fixed intensity of the average background field $\B = |eB|/m^2 = 10^4$. The eigenvalues corresponding to projections onto the four independent sub-spaces are displayed for comparison.}
    \label{fig:m_Delta}
\end{figure}

\begin{figure}[h!]
    \centering
    \,\,\includegraphics[scale=0.765]{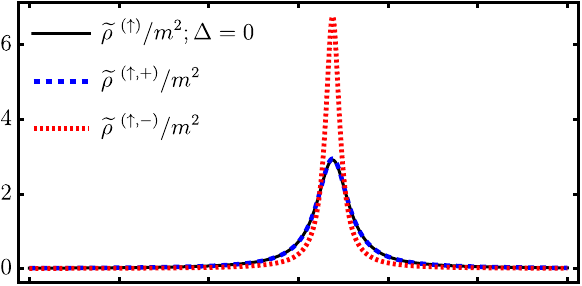}\nn\\
    \includegraphics[scale=0.775]{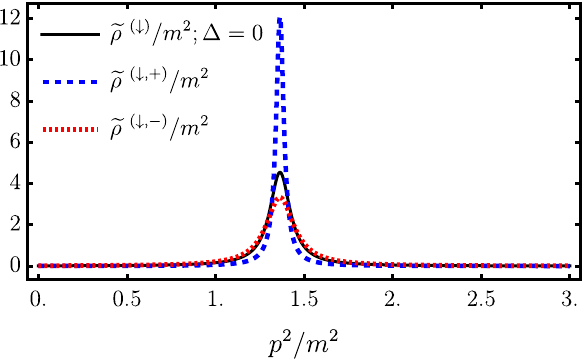}
    \caption{ }
    \label{fig:rho}
    \caption{The spectral density distributions for each of the four projections $\uparrow,\downarrow,\pm$, computed from Eq.~\eqref{eq:spectral}, as a function of the dimensionless momentum $p^2/m^2$. The dashed and dotted lines correspond to a noise auto-correlation $m\Delta = 10^{-2}$, while the solid line represents the noiseless limit $\Delta = 0$. The average background field is $\B = 10^5$ for all cases.}
\end{figure}

In Fig.~\ref{fig:m_Delta}, the imaginary part of the mass eigenvalues, corresponding to the Breit-Wigner resonances $\Gamma^{(\sigma,\lambda)}(\Delta)$ defined in Eqs.~\eqref{eq:mBupdown} for each of the four projections, are shown as a function of the magnetic noise autocorrelation $m\Delta$, for a constant average field value of $\B = 10^4$. In terms of the physical interpretation, these Breit-Wigner resonances proportional to $\text{Im}\Sigma(p,B,\Delta)$ lead to a small broadening in the peak of the Lorentzian spectral density distribution, as we discussed in Ref.~\cite{Ayala_mass_021}. As seen in Fig.~\ref{fig:m_Delta}, both parallel $(\uparrow,\pm)$ spin projections exhibit a linear dependence on $m\Delta$ with a negative slope. This effect is milder in the $(\uparrow,+)$ than in the $(\uparrow,-)$ polarization. In contrast,
the antiparallel spin polarizations $(\downarrow,\pm)$ display opposite behaviour, with $(\downarrow,-)$ showing a positive slope, while $(\downarrow,+)$ exhibits a negative one. Nevertheless, since the spectral broadening depends on the absolute value of these parameters, in all four polarizations the spectral width grows with the magnetic noise auto-correlation $\Delta$. As discussed in our previous work~\cite{Ayala_mass_021}, the spectral density corresponding to each projection is defined by the Lorentzian distributions
\bea
\widetilde{\rho}^{(\sigma,\lambda)}(p^2)=\frac{m_B^{(\sigma)}\Gamma^{(\sigma,\lambda)}/\pi}{\left[p^2-\left(m_B^{(\sigma)}\right)^2\right]^2+\left[m_B^{\sigma}\Gamma^{(\sigma,\lambda)}\right]^2},
\label{eq:spectral}
\eea
where the relative spectral width decays for very intense magnetic fields $\B \gg 1$ as
 $\frac{\Gamma^{(\sigma,\lambda)}}{m_B^{\sigma}} \sim \left[\ln\B\right]^{-1}$. The corresponding spectral density distributions, as computed from Eq.~\eqref{eq:spectral} for the four different projections, are displayed in Fig.~\ref{fig:rho}, where the spectral width due to the finite value of $\Gamma^{(\sigma,\lambda)}$ is clearly appreciated. Interestingly, this spectral broadening effect induced by the presence of the noise auto-correlation $\Delta > 0$ is different depending on the spin projection $\uparrow,\downarrow$, as well as the projection onto the subspaces $\mathcal{P}^{(\pm)}$. However, the physical magnetic mass representing the center of the spectral distribution, only depends on the spin projection, as expected from the usual Zeeman splitting effect due to the spin-magnetic field interaction.

\section{Discussion and Conclusions}

We studied the QED fermion propagator in a strongly magnetized medium with white-noise fluctuations. In a quasiparticle picture based on our previous results for the fermion and photon propagators in such media~\cite{Castano_Munoz_PhysRevD.107.096014,Castano_Munoz_PhysRevD.109.056007}, we computed the self-energy contribution due to radiative corrections at first-order in the electromagnetic fine structure constant $\alpha_\text{em}$. The presence of the background magnetic field breaks the Lorentz invariance, thus splitting the metric into two subspaces according to the directions parallel and perpendicular to the field, respectively. Accordingly, the eigenvalues of the magnetic mass operator obtained from the real part of the self-energy are different for the corresponding two spin projections $\uparrow, \downarrow$. Moreover, the presence of the magnetic fluctuations, whose strength is proportional to the noise auto-correlation $\Delta$, fully breaks the Lorentz invariance, thus leading to imaginary components in the mass operator eigenvalues that depend on the spin polarization as well as in the projection onto the $\mathcal{P}^{\pm} = \left( \mathbb{1} \pm \gamma^0 \right)/2$ subspaces corresponding to fermion ($+$) and anti-fermion ($-$) in their rest frame. The later imaginary contributions correspond to a further spectral broadening effect as we discussed in Ref.~\cite{Ayala_mass_021}, here caused by the magnetic noise that, nevertheless, does not renormalize the magnetic mass, in agreement with our previous studies~\cite{Castano_Munoz_PhysRevD.107.096014}.

\acknowledgements{ E.M. acknowledges financial support from Fondecyt 1230440. J.D.C.-Y. also acknowledges financial support from Fondecyt 3220087.}

\bibliography{SelfEnergyReplicas.bib}

\begin{thebibliography}{24}%
\makeatletter
\providecommand \@ifxundefined [1]{%
 \@ifx{#1\undefined}
}%
\providecommand \@ifnum [1]{%
 \ifnum #1\expandafter \@firstoftwo
 \else \expandafter \@secondoftwo
 \fi
}%
\providecommand \@ifx [1]{%
 \ifx #1\expandafter \@firstoftwo
 \else \expandafter \@secondoftwo
 \fi
}%
\providecommand \natexlab [1]{#1}%
\providecommand \enquote  [1]{``#1''}%
\providecommand \bibnamefont  [1]{#1}%
\providecommand \bibfnamefont [1]{#1}%
\providecommand \citenamefont [1]{#1}%
\providecommand \href@noop [0]{\@secondoftwo}%
\providecommand \href [0]{\begingroup \@sanitize@url \@href}%
\providecommand \@href[1]{\@@startlink{#1}\@@href}%
\providecommand \@@href[1]{\endgroup#1\@@endlink}%
\providecommand \@sanitize@url [0]{\catcode `\\12\catcode `\$12\catcode `\&12\catcode `\#12\catcode `\^12\catcode `\_12\catcode `\%12\relax}%
\providecommand \@@startlink[1]{}%
\providecommand \@@endlink[0]{}%
\providecommand \url  [0]{\begingroup\@sanitize@url \@url }%
\providecommand \@url [1]{\endgroup\@href {#1}{\urlprefix }}%
\providecommand \urlprefix  [0]{URL }%
\providecommand \Eprint [0]{\href }%
\providecommand \doibase [0]{http://dx.doi.org/}%
\providecommand \selectlanguage [0]{\@gobble}%
\providecommand \bibinfo  [0]{\@secondoftwo}%
\providecommand \bibfield  [0]{\@secondoftwo}%
\providecommand \translation [1]{[#1]}%
\providecommand \BibitemOpen [0]{}%
\providecommand \bibitemStop [0]{}%
\providecommand \bibitemNoStop [0]{.\EOS\space}%
\providecommand \EOS [0]{\spacefactor3000\relax}%
\providecommand \BibitemShut  [1]{\csname bibitem#1\endcsname}%
\let\auto@bib@innerbib\@empty
\bibitem [{\citenamefont {Alam}\ \emph {et~al.}(2021)\citenamefont {Alam}, \citenamefont {Roy}, \citenamefont {Ahmad},\ and\ \citenamefont {Chattopadhyay}}]{Alam_2021}%
  \BibitemOpen
  \bibfield  {author} {\bibinfo {author} {\bibfnamefont {Sk~Noor}\ \bibnamefont {Alam}}, \bibinfo {author} {\bibfnamefont {Victor}\ \bibnamefont {Roy}}, \bibinfo {author} {\bibfnamefont {Shakeel}\ \bibnamefont {Ahmad}}, \ and\ \bibinfo {author} {\bibfnamefont {Subhasis}\ \bibnamefont {Chattopadhyay}},\ }\bibfield  {title} {\enquote {\bibinfo {title} {Electromagnetic field fluctuation and its correlation with the participant plane in $\mathrm{Au}+\mathrm{Au}$ and isobaric collisions at $\sqrt{{s}_{NN}}=200\text{ }\text{ }\mathrm{GeV}$},}\ }\href {\doibase 10.1103/PhysRevD.104.114031} {\bibfield  {journal} {\bibinfo  {journal} {Phys. Rev. D}\ }\textbf {\bibinfo {volume} {104}},\ \bibinfo {pages} {114031} (\bibinfo {year} {2021})}\BibitemShut {NoStop}%
\bibitem [{\citenamefont {Ayala}\ \emph {et~al.}(2022)\citenamefont {Ayala}, \citenamefont {Casta{\~n}o-Yepes}, \citenamefont {Hern{\'a}ndez}, \citenamefont {Mizher}, \citenamefont {Tejeda-Yeomans},\ and\ \citenamefont {Zamora}}]{ayala2022anisotropic}%
  \BibitemOpen
  \bibfield  {author} {\bibinfo {author} {\bibfnamefont {Alejandro}\ \bibnamefont {Ayala}}, \bibinfo {author} {\bibfnamefont {Jorge~David}\ \bibnamefont {Casta{\~n}o-Yepes}}, \bibinfo {author} {\bibfnamefont {LA}~\bibnamefont {Hern{\'a}ndez}}, \bibinfo {author} {\bibfnamefont {Ana~Julia}\ \bibnamefont {Mizher}}, \bibinfo {author} {\bibfnamefont {Mar{\'\i}a~Elena}\ \bibnamefont {Tejeda-Yeomans}}, \ and\ \bibinfo {author} {\bibfnamefont {R}~\bibnamefont {Zamora}},\ }\bibfield  {title} {\enquote {\bibinfo {title} {{Anisotropic photon emission from gluon fusion and splitting in a strong magnetic background I: The two-gluon one-photon vertex}},}\ }\href {https://arxiv.org/abs/2209.09364} {\  (\bibinfo {year} {2022})},\ \Eprint {http://arxiv.org/abs/2209.09364} {arXiv:2209.09364 [hep-ph]} \BibitemShut {NoStop}%
\bibitem [{\citenamefont {{Inghirami, Gabriele}}\ \emph {et~al.}(2020)\citenamefont {{Inghirami, Gabriele}}, \citenamefont {{Mace, Mark}}, \citenamefont {{Hirono, Yuji}}, \citenamefont {{Del Zanna, Luca}}, \citenamefont {{Kharzeev, Dmitri E.}},\ and\ \citenamefont {{Bleicher, Marcus}}}]{Inghirami_2020}%
  \BibitemOpen
  \bibfield  {author} {\bibinfo {author} {\bibnamefont {{Inghirami, Gabriele}}}, \bibinfo {author} {\bibnamefont {{Mace, Mark}}}, \bibinfo {author} {\bibnamefont {{Hirono, Yuji}}}, \bibinfo {author} {\bibnamefont {{Del Zanna, Luca}}}, \bibinfo {author} {\bibnamefont {{Kharzeev, Dmitri E.}}}, \ and\ \bibinfo {author} {\bibnamefont {{Bleicher, Marcus}}},\ }\bibfield  {title} {\enquote {\bibinfo {title} {Magnetic fields in heavy ion collisions: flow and charge transport},}\ }\href {\doibase 10.1140/epjc/s10052-020-7847-4} {\bibfield  {journal} {\bibinfo  {journal} {Eur. Phys. J. C}\ }\textbf {\bibinfo {volume} {80}},\ \bibinfo {pages} {293} (\bibinfo {year} {2020})}\BibitemShut {NoStop}%
\bibitem [{\citenamefont {Ayala}\ \emph {et~al.}(2020{\natexlab{a}})\citenamefont {Ayala}, \citenamefont {Casta\~no Yepes}, \citenamefont {Dominguez~Jimenez}, \citenamefont {Salinas San~Mart\'\i{}n},\ and\ \citenamefont {Tejeda-Yeomans}}]{Ayala:2019jey}%
  \BibitemOpen
  \bibfield  {author} {\bibinfo {author} {\bibfnamefont {Alejandro}\ \bibnamefont {Ayala}}, \bibinfo {author} {\bibfnamefont {Jorge~David}\ \bibnamefont {Casta\~no Yepes}}, \bibinfo {author} {\bibfnamefont {Isabel}\ \bibnamefont {Dominguez~Jimenez}}, \bibinfo {author} {\bibfnamefont {Jordi}\ \bibnamefont {Salinas San~Mart\'\i{}n}}, \ and\ \bibinfo {author} {\bibfnamefont {Mar\'\i{}a~Elena}\ \bibnamefont {Tejeda-Yeomans}},\ }\bibfield  {title} {\enquote {\bibinfo {title} {{Centrality dependence of photon yield and elliptic flow from gluon fusion and splitting induced by magnetic fields in relativistic heavy-ion collisions}},}\ }\href {\doibase 10.1140/epja/s10050-020-00060-9} {\bibfield  {journal} {\bibinfo  {journal} {Eur. Phys. J. A}\ }\textbf {\bibinfo {volume} {56}},\ \bibinfo {pages} {53} (\bibinfo {year} {2020}{\natexlab{a}})}\BibitemShut {NoStop}%
\bibitem [{\citenamefont {Ayala}\ \emph {et~al.}(2017)\citenamefont {Ayala}, \citenamefont {Castano-Yepes}, \citenamefont {Dominguez}, \citenamefont {Hernandez}, \citenamefont {Hernandez-Ortiz},\ and\ \citenamefont {Tejeda-Yeomans}}]{Ayala:2017vex}%
  \BibitemOpen
  \bibfield  {author} {\bibinfo {author} {\bibfnamefont {Alejandro}\ \bibnamefont {Ayala}}, \bibinfo {author} {\bibfnamefont {Jorge~David}\ \bibnamefont {Castano-Yepes}}, \bibinfo {author} {\bibfnamefont {Cesareo~A.}\ \bibnamefont {Dominguez}}, \bibinfo {author} {\bibfnamefont {Luis~A.}\ \bibnamefont {Hernandez}}, \bibinfo {author} {\bibfnamefont {Saul}\ \bibnamefont {Hernandez-Ortiz}}, \ and\ \bibinfo {author} {\bibfnamefont {Maria~Elena}\ \bibnamefont {Tejeda-Yeomans}},\ }\bibfield  {title} {\enquote {\bibinfo {title} {{Prompt photon yield and elliptic flow from gluon fusion induced by magnetic fields in relativistic heavy-ion collisions}},}\ }\href {\doibase 10.1103/PhysRevD.96.014023} {\bibfield  {journal} {\bibinfo  {journal} {Phys. Rev. D}\ }\textbf {\bibinfo {volume} {96}},\ \bibinfo {pages} {014023} (\bibinfo {year} {2017})},\ \bibinfo {note} {[Erratum: Phys.Rev.D 96, 119901 (2017)]}\BibitemShut {NoStop}%
\bibitem [{\citenamefont {Busza}\ \emph {et~al.}(2018)\citenamefont {Busza}, \citenamefont {Rajagopal},\ and\ \citenamefont {van~der Schee}}]{Busza_2018}%
  \BibitemOpen
  \bibfield  {author} {\bibinfo {author} {\bibfnamefont {Wit}\ \bibnamefont {Busza}}, \bibinfo {author} {\bibfnamefont {Krishna}\ \bibnamefont {Rajagopal}}, \ and\ \bibinfo {author} {\bibfnamefont {Wilke}\ \bibnamefont {van~der Schee}},\ }\bibfield  {title} {\enquote {\bibinfo {title} {Heavy ion collisions: The big picture and the big questions},}\ }\href {\doibase 10.1146/annurev-nucl-101917-020852} {\bibfield  {journal} {\bibinfo  {journal} {Annual Review of Nuclear and Particle Science}\ }\textbf {\bibinfo {volume} {68}},\ \bibinfo {pages} {339--376} (\bibinfo {year} {2018})},\ \Eprint {http://arxiv.org/abs/https://doi.org/10.1146/annurev-nucl-101917-020852} {https://doi.org/10.1146/annurev-nucl-101917-020852} \BibitemShut {NoStop}%
\bibitem [{\citenamefont {Hattori}\ and\ \citenamefont {Satow}(2016)}]{Hattori_2016}%
  \BibitemOpen
  \bibfield  {author} {\bibinfo {author} {\bibfnamefont {Koichi}\ \bibnamefont {Hattori}}\ and\ \bibinfo {author} {\bibfnamefont {Daisuke}\ \bibnamefont {Satow}},\ }\bibfield  {title} {\enquote {\bibinfo {title} {Electrical conductivity of quark-gluon plasma in strong magnetic fields},}\ }\href {\doibase 10.1103/PhysRevD.94.114032} {\bibfield  {journal} {\bibinfo  {journal} {Phys. Rev. D}\ }\textbf {\bibinfo {volume} {94}},\ \bibinfo {pages} {114032} (\bibinfo {year} {2016})}\BibitemShut {NoStop}%
\bibitem [{\citenamefont {Hattori}\ and\ \citenamefont {Satow}(2018)}]{Hattori_2018}%
  \BibitemOpen
  \bibfield  {author} {\bibinfo {author} {\bibfnamefont {Koichi}\ \bibnamefont {Hattori}}\ and\ \bibinfo {author} {\bibfnamefont {Daisuke}\ \bibnamefont {Satow}},\ }\bibfield  {title} {\enquote {\bibinfo {title} {Gluon spectrum in a quark-gluon plasma under strong magnetic fields},}\ }\href {\doibase 10.1103/PhysRevD.97.014023} {\bibfield  {journal} {\bibinfo  {journal} {Phys. Rev. D}\ }\textbf {\bibinfo {volume} {97}},\ \bibinfo {pages} {014023} (\bibinfo {year} {2018})}\BibitemShut {NoStop}%
\bibitem [{\citenamefont {Buballa}(2005)}]{Buballa_2005}%
  \BibitemOpen
  \bibfield  {author} {\bibinfo {author} {\bibfnamefont {Michael}\ \bibnamefont {Buballa}},\ }\bibfield  {title} {\enquote {\bibinfo {title} {Njl-model analysis of dense quark matter},}\ }\href {\doibase https://doi.org/10.1016/j.physrep.2004.11.004} {\bibfield  {journal} {\bibinfo  {journal} {Physics Reports}\ }\textbf {\bibinfo {volume} {407}},\ \bibinfo {pages} {205--376} (\bibinfo {year} {2005})}\BibitemShut {NoStop}%
\bibitem [{\citenamefont {Schwinger}(1951)}]{Schwinger_1951}%
  \BibitemOpen
  \bibfield  {author} {\bibinfo {author} {\bibfnamefont {Julian}\ \bibnamefont {Schwinger}},\ }\bibfield  {title} {\enquote {\bibinfo {title} {On gauge invariance and vacuum polarization},}\ }\href {\doibase 10.1103/PhysRev.82.664} {\bibfield  {journal} {\bibinfo  {journal} {Phys. Rev.}\ }\textbf {\bibinfo {volume} {82}},\ \bibinfo {pages} {664--679} (\bibinfo {year} {1951})}\BibitemShut {NoStop}%
\bibitem [{\citenamefont {Dittrich}\ and\ \citenamefont {Reuter}(1985)}]{Dittrich_Reuter}%
  \BibitemOpen
  \bibfield  {author} {\bibinfo {author} {\bibfnamefont {W.}~\bibnamefont {Dittrich}}\ and\ \bibinfo {author} {\bibfnamefont {M.}~\bibnamefont {Reuter}},\ }\enquote {\bibinfo {title} {Effective {L}agrangians in {Q}uantum {E}lectrodynamics,lecture notes in physics},}\ \ (\bibinfo  {publisher} {Springer-Verlag},\ \bibinfo {address} {Berlin-Heidelberg},\ \bibinfo {year} {1985})\BibitemShut {NoStop}%
\bibitem [{\citenamefont {Dittrich}\ and\ \citenamefont {Gies}(2000)}]{Dittrich_Gies}%
  \BibitemOpen
  \bibfield  {author} {\bibinfo {author} {\bibfnamefont {W.}~\bibnamefont {Dittrich}}\ and\ \bibinfo {author} {\bibfnamefont {H.}~\bibnamefont {Gies}},\ }\enquote {\bibinfo {title} {Probing the quantum vacuum: Perturbative effective action approach in quantum electrodynamics and its application, springer tracts in modern physics},}\ \ (\bibinfo  {publisher} {Springer-Verlag},\ \bibinfo {address} {Berlin-Heidelberg},\ \bibinfo {year} {2000})\BibitemShut {NoStop}%
\bibitem [{\citenamefont {Hattori}\ and\ \citenamefont {Itakura}(2013)}]{Hattori_2013}%
  \BibitemOpen
  \bibfield  {author} {\bibinfo {author} {\bibfnamefont {Koichi}\ \bibnamefont {Hattori}}\ and\ \bibinfo {author} {\bibfnamefont {Kazunori}\ \bibnamefont {Itakura}},\ }\bibfield  {title} {\enquote {\bibinfo {title} {Vacuum birefringence in strong magnetic fields: (i) photon polarization tensor with all the landau levels},}\ }\href {\doibase https://doi.org/10.1016/j.aop.2012.11.010} {\bibfield  {journal} {\bibinfo  {journal} {Annals of Physics}\ }\textbf {\bibinfo {volume} {330}},\ \bibinfo {pages} {23--54} (\bibinfo {year} {2013})}\BibitemShut {NoStop}%
\bibitem [{\citenamefont {Ayala}\ \emph {et~al.}(2020{\natexlab{b}})\citenamefont {Ayala}, \citenamefont {Casta\~no Yepes}, \citenamefont {Loewe},\ and\ \citenamefont {Mu\~noz}}]{Ayala_Pol_020}%
  \BibitemOpen
  \bibfield  {author} {\bibinfo {author} {\bibfnamefont {Alejandro}\ \bibnamefont {Ayala}}, \bibinfo {author} {\bibfnamefont {Jorge~David}\ \bibnamefont {Casta\~no Yepes}}, \bibinfo {author} {\bibfnamefont {M.}~\bibnamefont {Loewe}}, \ and\ \bibinfo {author} {\bibfnamefont {Enrique}\ \bibnamefont {Mu\~noz}},\ }\bibfield  {title} {\enquote {\bibinfo {title} {Gluon polarization tensor in a magnetized medium: Analytic approach starting from the sum over landau levels},}\ }\href {\doibase 10.1103/PhysRevD.101.036016} {\bibfield  {journal} {\bibinfo  {journal} {Phys. Rev. D}\ }\textbf {\bibinfo {volume} {101}},\ \bibinfo {pages} {036016} (\bibinfo {year} {2020}{\natexlab{b}})}\BibitemShut {NoStop}%
\bibitem [{\citenamefont {Tsai}(1974)}]{Tsai_PhysRevD.10.1342}%
  \BibitemOpen
  \bibfield  {author} {\bibinfo {author} {\bibfnamefont {Wu-yang}\ \bibnamefont {Tsai}},\ }\bibfield  {title} {\enquote {\bibinfo {title} {Modified electron propagation function in strong magnetic fields},}\ }\href {\doibase 10.1103/PhysRevD.10.1342} {\bibfield  {journal} {\bibinfo  {journal} {Phys. Rev. D}\ }\textbf {\bibinfo {volume} {10}},\ \bibinfo {pages} {1342--1345} (\bibinfo {year} {1974})}\BibitemShut {NoStop}%
\bibitem [{\citenamefont {Ayala}\ \emph {et~al.}(2021)\citenamefont {Ayala}, \citenamefont {Casta\~no Yepes}, \citenamefont {Loewe},\ and\ \citenamefont {Mu\~noz}}]{Ayala_mass_021}%
  \BibitemOpen
  \bibfield  {author} {\bibinfo {author} {\bibfnamefont {Alejandro}\ \bibnamefont {Ayala}}, \bibinfo {author} {\bibfnamefont {Jorge~David}\ \bibnamefont {Casta\~no Yepes}}, \bibinfo {author} {\bibfnamefont {M.}~\bibnamefont {Loewe}}, \ and\ \bibinfo {author} {\bibfnamefont {Enrique}\ \bibnamefont {Mu\~noz}},\ }\bibfield  {title} {\enquote {\bibinfo {title} {Fermion mass and width in qed in a magnetic field},}\ }\href {\doibase 10.1103/PhysRevD.104.016006} {\bibfield  {journal} {\bibinfo  {journal} {Phys. Rev. D}\ }\textbf {\bibinfo {volume} {104}},\ \bibinfo {pages} {016006} (\bibinfo {year} {2021})}\BibitemShut {NoStop}%
\bibitem [{\citenamefont {Machet}(2016)}]{Machet_10.1142/S0217751X16500718}%
  \BibitemOpen
  \bibfield  {author} {\bibinfo {author} {\bibfnamefont {B.}~\bibnamefont {Machet}},\ }\bibfield  {title} {\enquote {\bibinfo {title} {The 1-loop self-energy of an electron in a strong external magnetic field revisited},}\ }\href {\doibase 10.1142/S0217751X16500718} {\bibfield  {journal} {\bibinfo  {journal} {International Journal of Modern Physics A}\ }\textbf {\bibinfo {volume} {31}},\ \bibinfo {pages} {1650071} (\bibinfo {year} {2016})},\ \Eprint {http://arxiv.org/abs/https://doi.org/10.1142/S0217751X16500718} {https://doi.org/10.1142/S0217751X16500718} \BibitemShut {NoStop}%
\bibitem [{\citenamefont {JANCOVICI}(1969)}]{Jancovici_PhysRev.187.2275}%
  \BibitemOpen
  \bibfield  {author} {\bibinfo {author} {\bibfnamefont {B.}~\bibnamefont {JANCOVICI}},\ }\bibfield  {title} {\enquote {\bibinfo {title} {Radiative correction to the ground-state energy of an electron in an intense magnetic field},}\ }\href {\doibase 10.1103/PhysRev.187.2275} {\bibfield  {journal} {\bibinfo  {journal} {Phys. Rev.}\ }\textbf {\bibinfo {volume} {187}},\ \bibinfo {pages} {2275--2276} (\bibinfo {year} {1969})}\BibitemShut {NoStop}%
\bibitem [{\citenamefont {Casta\~no Yepes}\ \emph {et~al.}(2023)\citenamefont {Casta\~no Yepes}, \citenamefont {Loewe}, \citenamefont {Mu\~noz}, \citenamefont {Rojas},\ and\ \citenamefont {Zamora}}]{Castano_Munoz_PhysRevD.107.096014}%
  \BibitemOpen
  \bibfield  {author} {\bibinfo {author} {\bibfnamefont {Jorge~David}\ \bibnamefont {Casta\~no Yepes}}, \bibinfo {author} {\bibfnamefont {Marcelo}\ \bibnamefont {Loewe}}, \bibinfo {author} {\bibfnamefont {Enrique}\ \bibnamefont {Mu\~noz}}, \bibinfo {author} {\bibfnamefont {Juan~Crist\'obal}\ \bibnamefont {Rojas}}, \ and\ \bibinfo {author} {\bibfnamefont {Renato}\ \bibnamefont {Zamora}},\ }\bibfield  {title} {\enquote {\bibinfo {title} {Qed fermions in a noisy magnetic field background},}\ }\href {\doibase 10.1103/PhysRevD.107.096014} {\bibfield  {journal} {\bibinfo  {journal} {Phys. Rev. D}\ }\textbf {\bibinfo {volume} {107}},\ \bibinfo {pages} {096014} (\bibinfo {year} {2023})}\BibitemShut {NoStop}%
\bibitem [{\citenamefont {M{\'e}zard}\ and\ \citenamefont {Parisi}(1991)}]{Parisi_Mezard_1991}%
  \BibitemOpen
  \bibfield  {author} {\bibinfo {author} {\bibfnamefont {Marc}\ \bibnamefont {M{\'e}zard}}\ and\ \bibinfo {author} {\bibfnamefont {Giorgio}\ \bibnamefont {Parisi}},\ }\bibfield  {title} {\enquote {\bibinfo {title} {Replica field theory for random manifolds},}\ }\href {\doibase https://doi.org/10.1051/jp1:1991171} {\bibfield  {journal} {\bibinfo  {journal} {Journal de Physique I}\ }\textbf {\bibinfo {volume} {1}},\ \bibinfo {pages} {809--836} (\bibinfo {year} {1991})}\BibitemShut {NoStop}%
\bibitem [{\citenamefont {Kardar}\ \emph {et~al.}(1986)\citenamefont {Kardar}, \citenamefont {Parisi},\ and\ \citenamefont {Zhang}}]{Kardar_Parisi_1986}%
  \BibitemOpen
  \bibfield  {author} {\bibinfo {author} {\bibfnamefont {M.}~\bibnamefont {Kardar}}, \bibinfo {author} {\bibfnamefont {G.}~\bibnamefont {Parisi}}, \ and\ \bibinfo {author} {\bibfnamefont {Y.-C.}\ \bibnamefont {Zhang}},\ }\bibfield  {title} {\enquote {\bibinfo {title} {{Dynamic scaling of growing interfaces}},}\ }\href {\doibase https://doi.org/10.1103/PhysRevLett.56.889} {\bibfield  {journal} {\bibinfo  {journal} {Phys. Rev. Lett.}\ }\textbf {\bibinfo {volume} {56}},\ \bibinfo {pages} {889--892} (\bibinfo {year} {1986})}\BibitemShut {NoStop}%
\bibitem [{\citenamefont {Casta\~no Yepes}\ and\ \citenamefont {Mu\~noz}(2024)}]{Castano_Munoz_PhysRevD.109.056007}%
  \BibitemOpen
  \bibfield  {author} {\bibinfo {author} {\bibfnamefont {Jorge~David}\ \bibnamefont {Casta\~no Yepes}}\ and\ \bibinfo {author} {\bibfnamefont {Enrique}\ \bibnamefont {Mu\~noz}},\ }\bibfield  {title} {\enquote {\bibinfo {title} {{Exploring magnetic fluctuation effects in QED gauge fields: Implications for mass generation}},}\ }\href {\doibase 10.1103/PhysRevD.109.056007} {\bibfield  {journal} {\bibinfo  {journal} {Phys. Rev. D}\ }\textbf {\bibinfo {volume} {109}},\ \bibinfo {pages} {056007} (\bibinfo {year} {2024})}\BibitemShut {NoStop}%
\bibitem [{\citenamefont {Fukushima}(2011)}]{Fukushima_PhysRevD.83.111501}%
  \BibitemOpen
  \bibfield  {author} {\bibinfo {author} {\bibfnamefont {Kenji}\ \bibnamefont {Fukushima}},\ }\bibfield  {title} {\enquote {\bibinfo {title} {Magnetic-field induced screening effect and collective excitations},}\ }\href {\doibase 10.1103/PhysRevD.83.111501} {\bibfield  {journal} {\bibinfo  {journal} {Phys. Rev. D}\ }\textbf {\bibinfo {volume} {83}},\ \bibinfo {pages} {111501} (\bibinfo {year} {2011})}\BibitemShut {NoStop}%
\bibitem [{\citenamefont {Bandyopadhyay}\ \emph {et~al.}(2016)\citenamefont {Bandyopadhyay}, \citenamefont {Islam},\ and\ \citenamefont {Mustafa}}]{Bandyopadhyay_PhysRevD.94.114034}%
  \BibitemOpen
  \bibfield  {author} {\bibinfo {author} {\bibfnamefont {Aritra}\ \bibnamefont {Bandyopadhyay}}, \bibinfo {author} {\bibfnamefont {Chowdhury~Aminul}\ \bibnamefont {Islam}}, \ and\ \bibinfo {author} {\bibfnamefont {Munshi~G.}\ \bibnamefont {Mustafa}},\ }\bibfield  {title} {\enquote {\bibinfo {title} {Electromagnetic spectral properties and debye screening of a strongly magnetized hot medium},}\ }\href {\doibase 10.1103/PhysRevD.94.114034} {\bibfield  {journal} {\bibinfo  {journal} {Phys. Rev. D}\ }\textbf {\bibinfo {volume} {94}},\ \bibinfo {pages} {114034} (\bibinfo {year} {2016})}\BibitemShut {NoStop}%
\end{thebibliography}%

\appendix

\section{Computation of $-\ii\Sigma_{0}(p,B)$}\label{Ap:Sigma00}
From Eqs.~\eqref{eq:S0def},~\eqref{eq:D0def} and~\eqref{eq:Sigma00def}, we get
{\small
\bea
-\ii\Sigma_{0}(p,B)=2 e^2\int\frac{d^4k}{(2\pi)^4}\frac{e^{-\kt^2/\qB}\gamma^\mu(\slashed{k}_\parallel+m)\Op{\uparrow}\gamma_\mu}{(\kp^2-m^2+\ii\epsilon)((k-p)^2+\ii\epsilon)},\nn\\
\eea
}
so that the tensor structure can be spitted as:
\bea
&&\gamma^\mu(\slashed{k}_\parallel+m)\Op{\uparrow}\gamma_\mu\nn\\
&=&\frac{1}{2}\Big[\gm\slashed{k}_\parallel\gamma_\mu-\ii\,\sign\gm\slashed{k}_\parallel\gamma^1\gamma^2\gamma_\mu+m\gm\gamma_\mu\nn\\
&-&\ii\,\sign m\gm\gamma^1\gamma^2\gamma_\mu\Big]\nn\\
&=&\frac{1}{2}\Big[-2\slashed{k}_\parallel+2\ii\,\sign\gamma^2\gamma^1\slashed{k}_\parallel+4m\Big]\nn\\
&=&2\left[m-\slashed{k}_\parallel\Op{\downarrow}\right].
\eea

On the other hand, by a Schwinger parametrization (in the Feynman time-ordered prescription $\epsilon\rightarrow 0^{+}$),
\bea
\frac{1}{A + \ii \epsilon}=-\ii\int_0^\infty d\tau\, e^{\ii ( A + \ii \epsilon)\tau },
\eea
the expression takes the form
 {\small
 \bea
&&-\ii\Sigma_{0}(p,B)\nn\\
 &=&4 e^2(-\ii)^2\int_0^{\infty} d\tau_1\int_0^{\infty}d\tau_2\int\frac{d^4k}{(2\pi)^4}\left[m-\slashed{k}_\parallel\Op{\downarrow}\right]\nn\\
 &\times&\exp\left\{-\frac{\kt^2}{\qB}+\ii\tau_1\left[(k-p)^2+\ii\epsilon\right]+\ii \tau_2(\kp^2-m^2+\ii\epsilon)\right\}.\nn\\
 \eea
 }

The factor in the exponential can be rearranged as:
\bea
&&-\frac{\kt^2}{\qB}+\ii\tau_1\left[(k-p)^2+\ii\epsilon\right]+\ii \tau_2(\kp^2-m^2+\ii\epsilon)\nn\\
&=&-\left(\frac{1}{\qB}+\ii\tau_1\right)\left(\kt^2-\frac{2\ii\tau_1 k_\perp\cdot p_\perp}{1/\qB+\ii\tau_1}\right)-\ii \tau_1\pt^2\nn\\
&+&\ii(\tau_1+\tau_2)\left(\kp^2-\frac{2\tau_1 \kp\cdot\pp}{\tau_1+\tau_2}\right)\nn\\
&+&\ii\tau_1\pp^2-\ii \tau_2 m^2-\epsilon(\tau_1+\tau_2),\nn\\
\eea
so that, by performing the following change of variables
\bea
\ell_\parallel^\mu&\equiv&\kp^\mu-\frac{\tau_1}{\tau_1+\tau_2}\pp^\mu\nn\\
\ell_\perp&\equiv&k_\perp^\mu-\frac{\ii\qB\tau_1}{1+\ii\qB\tau_1}p_\perp^\mu,
\label{eq:lpara_lperp}
\eea
we get
\begin{widetext}
{\small
\bea
-\ii\Sigma_{0}(p,B)&=&-4 e^2\int d^2\tau\int\frac{d^4\ell}{(2\pi)^4}\left[m-\left(\slashed{\ell}_\parallel+\frac{\tau_1}{\tau_1+\tau_2}\slashed{p}_\parallel\right)\Op{\downarrow}\right]\nn\\
&&\qquad\qquad\qquad\qquad\quad\times\exp\left\{\ii(\tau_1+\tau_2)\ell_\parallel^2-\frac{1+\ii\qB\tau_1}{\qB}\ell_\perp^2+\ii\tau_2\left(\frac{\tau_1 \pp^2}{\tau_1+\tau_2}-m^2\right)-\frac{\ii \tau_1\pt^2}{1+\ii\qB \tau_1}-\epsilon(\tau_1+\tau_2)\right\}.\nn\\
\eea
}  
\end{widetext}

The Gaussian integration over the momenta are calculated as follows,
\begin{subequations}
\bea
\int\frac{d^2\ell_\parallel}{(2\pi)^2}\exp\left[\ii(\tau_1+\tau_2)\ell_\parallel^2\right]&=& \ii \int\frac{d^2\ell_{E}}{(2\pi)^2}e^{-\ii(\tau_1+\tau_2)\ell_{E}^2}\nn\\
&=& 
\frac{\ii}{(2\pi)^2}\frac{\pi}{\ii(\tau_1+\tau_2)}\nn\\
&=& 
\frac{1}{(2\pi)^2}\frac{\pi}{(\tau_1+\tau_2)}
\label{eq:gaussianint_perp}
\eea
\bea
     \int\frac{d^2\ell_\perp}{(2\pi)^2}\exp\left[-\frac{1+\ii\qB\tau_1}{\qB}\ell_\perp^2\right]=\frac{1}{(2\pi)^2}\frac{\pi\qB}{1+\ii\qB\tau_1}
     \label{eq:gaussianint_para},\nn\\
\eea
\label{eq:gaussianint}
\end{subequations}
where in the first equation we first performed a Wick rotation $\ell^0\rightarrow \ii \ell^0_E$ to render the integration coordinates to the Euclidean metric.
After this procedure, we are left with the expression
{\small
\bea
&&-\ii\Sigma_{0}(p,B)=\frac{\alpha_\text{em}\qB}{\pi}\int d^2\tau \frac{m-\frac{\tau_1}{\tau_1+\tau_2}\slashed{p}_\parallel\Op{\downarrow}}{(1+\ii\qB\tau_1)(\tau_1+\tau_2)}\nn\\
&\times&\exp\left[\ii\tau_2\left(\frac{\tau_1 \pp^2}{\tau_1+\tau_2}-m^2\right)-\frac{\ii \tau_1\pt^2}{1+\ii\qB \tau_1}-\epsilon(\tau_1+\tau_2)\right].\nn\\
\eea
}
It is convenient to introduce the change of variables
\bea
\tau_1=\frac{sy}{m^2}~\text{and}~\tau_2=\frac{s(1-y)}{m^2}\to\left|\frac{\partial(\tau_1,\tau_2)}{\partial(s,y)}\right|=\frac{s}{m^4},\nn\\
\label{eq:tau1tau2}
\eea
so that in the dimensionless variables $\mathcal{B}=\qB/m^2$, $\rho_{\parallel,\perp}=p_{\parallel,\perp}/m$
\begin{widetext}
\bea
-\ii\Sigma_{0}(p,B)=\frac{\alpha_\text{em}\B m}{\pi}\int_0^\infty ds\int_0^1 dy\frac{1-y\slashed{\rho}_\parallel\Op{\downarrow}}{1+\ii\B sy}\exp\left[\ii s(1-y)\left(y\rp^2-1\right)-\frac{\ii sy\rt^2}{1+\ii\B sy}-\epsilon s\right].\nn\\
\label{eq_Sig00}
\eea
\end{widetext}
In order to obtain the expression for the magnetic mass operator in the noiseless limit $\Delta = 0$ as stated in the main text, we evaluate this integral at the condition $\slashed{p}_{\parallel} = m$, $\mathbf{p}_{\perp} = 0$, as follows
\begin{widetext}
\bea
\left.\Sigma_{0}(p,B)\right|_{\slashed{p}_{\parallel} = m,\mathbf{p}_{\perp}=0}&=&\frac{\ii\alpha_\text{em}\B m}{\pi}\int_0^\infty ds\int_0^1 dy\frac{1-y\Op{\downarrow}}{1+\ii\B sy}\exp\left(-\ii s(1-y)^2-\epsilon s\right)\nn\\
&=&\frac{\ii\alpha_\text{em}\B m}{\pi}\left[\left(\int_0^\infty ds\int_0^1 dy\frac{e^{-\ii s (1-y)^2-\epsilon s}}{1+\ii\B sy}\right)\Op{\uparrow}+\left(\int_0^\infty ds\int_0^1 dy\frac{(1-y)e^{-\ii s (1-y)^2-\epsilon s}}{1+\ii\B sy}\right)\Op{\downarrow}\right]\nn\\
&=&  M_{B0}^{(\uparrow)}\Op{\uparrow} + M_{B0}^{(\downarrow)}\Op{\downarrow} 
\eea
\end{widetext}
where we used the completeness relation for the spin projectors
\bea
\Op{\uparrow}+\Op{\downarrow}=\mathbb{1},
\eea
and defined the corresponding fermion magnetic mass eigenvalues $M_{B0}^{\uparrow,\downarrow}$ for the two spin projections $\uparrow,\downarrow$, respectively. In order to explicitly compute such noiseless mass eigenvalues for the intense magnetic field approximation $\mathcal{B} \gg 1$, the integration region is restricted by a lower cutoff~\cite{Ayala_mass_021} $\sim \mathcal{B}^{-1}$, as follows (for $\epsilon\rightarrow 0^{+}$)
\begin{widetext}
\bea
&&\int_0^\infty ds\int_0^1 dy\frac{e^{-\ii s (1-y)^2 -\epsilon s}}{1+\ii\B sy}\rightarrow \int_{\B^{-1}}^\infty ds\int_{\B^{-1}}^1 dy \frac{e^{-\ii s (1-y)^2 -\epsilon s}}{1+\ii\B sy} = \int_{\B^{-1}}^\infty ds\int_{\B^{-1}}^1 dy e^{-\ii s (1-y)^2 -\epsilon s} \left( -\frac{\ii}{\mathcal{B}s y} - \sum_{n=1}^{\infty}\left(\frac{\ii}{\mathcal{B}sy}\right)^{n+1} \right)\nn\\
&=&\int_{\B^{-1}}^1 \frac{dy}{\ii\B y}\left\{\Gamma\left[0;\frac{\ii(1-y)^2}{\B}\right]+\sum_{n=1}^{\infty}\left(\frac{\ii}{\mathcal{B}} \right)^{n}\frac{\left[\ii (1-y)^2 \right]^{n}}{y^{n}}\Gamma\left[-n;\frac{\ii(1-y)^2}{\B}\right]\right\},
\label{eq_Int1}
\eea
\end{widetext}
Let us consider the series representation of the incomplete Gamma function, defined by
\bea
\Gamma(0,z) = -\gamma_e - \ln (z) - \sum_{k=1}^{\infty}\frac{(-z)^k}{k\cdot k!},
\eea
where $\gamma_\text{e}$ is the Euler-Mascheroni constant,
and for $n \in \mathbb{Z}^{+}$ by
\bea
\Gamma(-n,z) &=& \frac{(-1)^{n}}{n!}\Gamma(0,z) + \frac{z^{-n}f_n (z) e^{-z}}{n!},
\eea
with the $n-1$ degree polynomials
\be
f_n(z) = \sum_{k=0}^{n-1}(-1)^k (n-k-1)! z^{k}.
\ee

Then, it is convenient to re-organize the infinite series in the integrand of Eq.~\eqref{eq_Int1} in order to group the leading logarithmic contributions in the $\Gamma\left[0;\frac{\ii(1-y)^2}{\B}\right]$ function as follows
\begin{widetext}
\bea
&&\frac{1}{\ii\B y}\left\{\Gamma\left[0;\frac{\ii(1-y)^2}{\B}\right]
+\sum_{n=1}^{\infty}\left(\frac{\ii}{\mathcal{B}} \right)^{n}\frac{\left[\ii (1-y)^2 \right]^{n}}{y^{n}}\Gamma\left[-n;\frac{\ii(1-y)^2}{\B}\right]\right\} = 
\frac{1}{\ii\B y}\left\{ 1 + \sum_{n=1}^{\infty}\frac{1}{n!}\left(\frac{(1-y)^2}{y\B}\right)^n  \right\}
\Gamma\left[0;\frac{\ii(1-y)^2}{\B}\right]\nn\\
&+& \frac{e^{-\ii \frac{(1-y)^2}{\B y}}}{\ii\B y} \sum_{n=1}^{\infty}\frac{(-\ii\B y)^{
-n}}{n!}f_n\left( \ii (1-y)^2 /\B \right)\nn\\
&=& \frac{e^{\frac{(1-y)^2}{\B y}}}{\ii\B y} \Gamma\left[0;\frac{\ii(1-y)^2}{\B}\right] + \frac{e^{-\ii \frac{(1-y)^2}{\B y}}}{\ii\B y} \sum_{n=1}^{\infty}\frac{(-\ii\B y)^{
-n}}{n!}f_n\left( \ii (1-y)^2 /\B \right) \approx \frac{\ii\left(  \gamma_e - \ln\B + \ln\left[ \ii(1-y)^2\right]\right)}{y\B} + O(\B^{-2})
\eea
\end{widetext}

Then (with $\epsilon\to0^{+}$), we have that at leading order in $\mathcal{B}\gg1$
\begin{widetext}
\bea
\int_{\mathcal{B}^{-1}}^\infty ds\int_{\mathcal{B}^{-1}}^1 dy\frac{e^{-\ii s (1-y)^2 -\epsilon s}}{1+\ii\B sy}&\approx&
\int_{\mathcal{B}^{-1}}^1 dy\frac{\ii\left(  \gamma_e - \ln\B + \ln\left[ \ii(1-y)^2\right]\right)}{y\B} + O(\B^{-2})
\nn\\
&\approx&\frac{\ii}{\B}\left\{\left[\gamma_e + \ln\left(\frac{\ii(1-\B)^2}{\B^3}\right)\right]\ln(\B)-2\text{Li}_2\left(\frac{\B-1}{\B}\right)\right\}+O\left(\B^{-2}\right)\nn\\
&\approx& -\ii \B^{-1} \ln^2\B + \ii \left(\gamma_e + \ii\frac{\pi}{2}\right)\B^{-1}\ln\B - \ii \frac{\pi^2}{3}\B^{-1} + O(\B^{-2}) 
\eea
\end{widetext}

Applying the same series expansion to the second term, we have
\begin{widetext}
\bea
\int_{\mathcal{B}^{-1}}^\infty ds\int_{\mathcal{B}^{-1}}^1 dy\frac{(1-y)e^{-\ii s (1-y)^2 -\epsilon s}}{1+\ii\B sy}
&\approx& \int_{\mathcal{B}^{-1}}^1 dy (1 - y)\frac{\ii\left(  \gamma_e - \ln\B + \ln\left[ \ii(1-y)^2\right]\right)}{y\B} + O(\B^{-2})\\
&\approx& -\ii\left( \frac{(\gamma_e - 2)}{\B}- \gamma_e \B^{-1}\ln\B + \frac{(1 - \ln\B - \B^{-1})}{\B}\ln\left[ \frac{\ii(\B-1)^2}{\B^3} \right]\right)\nn\\ 
&-&2\ii\B^{-1}\text{Li}_2\left(\frac{\B-1}{\B}\right)   + O(\B^{-2})\nn\\
&\approx& -\ii \B^{-1} \ln^2\B + \ii \left(1 + \gamma_e + \ii\frac{\pi}{2}\right)\B^{-1}\ln\B + \ii\left( 2 - \gamma_e - \frac{\pi^2}{3} - \ii\frac{\pi}{2}  \right)\B^{-1} + O(\B^{-2})\nn
\eea
\end{widetext}

Therefore, we conclude that in the noiseless limit $\Delta = 0$, the magnetic mass eigenvalues for each spin projection are given by the expressions
\bea
M_{B0}^{(\uparrow)} &=& \frac{\alpha_\text{em}m}{\pi}\left[   \ln^2\B - \left( \gamma_e + \ii\frac{\pi}{2}\right)\ln\B+ \frac{\pi^2}{3}  \right]
+ O(\B^{-1})\nn\\
M_{B0}^{(\downarrow)} &=& \frac{\alpha_\text{em}m}{\pi}\left[   \ln^2\B - \left(1 + \gamma_e + \ii\frac{\pi}{2}\right)\ln\B\right.\nn\\
&&\left.-\left( 2 - \gamma_e - \frac{\pi^2}{3} - \ii\frac{\pi}{2}  \right)  \right]
+ O(\B^{-1})
\eea

\twocolumngrid
\section{Calculation of $-\ii\Sigma_{\Delta}(p)$}\label{Ap:SigmaDelta}

From Eqs.~\eqref{eq:SDdef},~\eqref{eq:D0def}  and~\eqref{eq:SigmaD0def}, and given that
\begin{subequations}
    \bea
    \gamma^\mu(\slashed{k}_\parallel+m)\Op{\uparrow}\gamma_\mu=2\left[m-\slashed{k}_\parallel\Op{\downarrow}\right],
    \eea
    \bea
    \gamma^\mu\gamma^3\Op{\uparrow}\gamma_\mu&=&-2\gamma^3\Op{\downarrow}
    \eea
    \bea
    \gamma^\mu\ii\gamma^1\gamma^2(\slashed{k}_\parallel+m)\gamma_\mu=2\slashed{k}_\parallel\ii\gamma^1\gamma^2,
    \eea
\end{subequations}
we get
\bea
-\ii\Sigma_{\Delta}(p)=\sum_{i=1}^3\left[-\ii\Sigma_{\Delta}^{(i)}(p)\right].
\eea
where
{\small
\begin{subequations}
    \bea
    -\ii\Sigma_{\Delta}^{(1)}(p)=-4\alpha_\text{em}\Delta\qB\int\frac{d^4k}{(2\pi)^4}\frac{\Theta_1(k)\left(m-\slashed{k}_\parallel\Op{\downarrow}\right)}{(k-p)^2+\ii\epsilon},\nn\\
    \eea
    \bea
    -\ii\Sigma_{\Delta}^{(2)}(p)=-4\alpha_\text{em}\Delta\qB\int\frac{d^4k}{(2\pi)^4}\frac{\Theta_2(k)\gamma^3\Op{\downarrow}}{(k-p)^2+\ii\epsilon}
    \eea
    \bea
    -\ii\Sigma_{\Delta}^{(3)}(p)=-4\alpha_\text{em}\Delta\qB\sign\int\frac{d^4k}{(2\pi)^4}\frac{\Theta_3(k)\slashed{k}_\parallel\ii\gamma^1\gamma^2}{(k-p)^2+\ii\epsilon}.\nn\\
    \eea
\end{subequations}
}

Let us start with $-\ii\Sigma_{\Delta}^{(1)}(p)$, which from Eq.~\eqref{Theta1} is
\begin{widetext}
 \bea
-\ii\Sigma_{\Delta}^{(1)}(p)=-12\alpha_\text{em}\Delta\qB\int\frac{d^4k}{(2\pi)^4}\frac{(\kp^2+m^2)e^{-2\kt^2/\qB}}{\left[(\kp^2-m^2)^2\sqrt{k_0^2-m^2}\right]\left[(k-p)^2+\ii\epsilon\right]}\left(m-\slashed{k}_\parallel\Op{\downarrow}\right).
\eea   
\end{widetext}

After a Schwinger parametrization in the photon's propagator and by defining the momenta shift
\bea
\ell_\perp=k_\perp^\mu-\frac{\ii\qB }{2+\ii\qB\tau}p_\perp^\mu,
\label{Lperp_2}
\eea
the integral takes the form
{\small
\bea
-\ii\Sigma_{\Delta}^{(1)}(p)&=&12\ii\alpha_\text{em}\Delta\qB\int_0^\infty d\tau\exp\left[-\frac{2\ii\tau\pt^2}{2+\ii\tau\qB}-\epsilon\tau\right]\nn\\
&\times&\int\frac{d^2\ell_\perp}{(2\pi)^2}\exp\left[-\left(\frac{2+\ii\qB\tau}{\qB}\right)\ell_\perp^2\right]\nn\\
&\times&\int\frac{d^2k_\parallel}{(2\pi)^2}\frac{\left(k_\parallel+m^2\right)e^{\ii\tau(\kp-\pp)^2}}{(\kp^2-m^2)^2\sqrt{k_0^2-m^2}}\left(m-\slashed{k}_\parallel\Op{\downarrow}\right).\nn\\
\eea
}

By using Eq.~\eqref{eq:gaussianint_perp}, and defining the parallel momenta shift:
\bea
\ell_\parallel^\mu&=&\kp^\mu-\pp^\mu,
\label{Lpara_2}
\eea
{\small
\bea
-\ii\Sigma_{\Delta}^{(1)}(p)&=&\frac{3\ii\alpha_\text{em}\Delta\qB^2}{\pi}\nn\\
&\times&\int_0^\infty d\tau\frac{\exp\left[-\frac{2\ii\tau\pt^2}{2+\ii\tau\qB}-\epsilon\tau\right]}{2+\ii\qB\tau}\mathcal{I}_1(\tau,p_0,p_3),\nn\\
\eea
}
where
{\small
\bea
&&\mathcal{I}_1(\tau,p_0,p_3)\nn\\
&\equiv&\int\frac{d^2\ell_\parallel}{(2\pi)^2}\frac{\left[\left(\ell_\parallel+p_\parallel\right)^2+m^2\right]\left[m-\left(\slashed{\ell}_\parallel+\slashed{p}_\parallel\right)\Op{\downarrow}\right]e^{\ii\tau\ell_\parallel^2}}{\left[\left(\ell_\parallel+p_\parallel\right)^2-m^2\right]^2\sqrt{\left(\ell_0+p_0\right)^2-m^2}}.\nn\\
\eea
}

We are interested on the limits $p_0\to m$ and $\mathbf{p}\to0$, so that
{\small
\bea
\lim_{\substack{p_0\to m\\\mathbf{p}\to 0}}\left[-\ii\Sigma_{\Delta}^{(1)}(p)\right]=\frac{3\ii\alpha_\text{em}\Delta\qB^2}{\pi}\int_0^\infty d\tau\frac{\mathcal{I}_1(\tau,m,0)e^{-\epsilon\tau}}{2+\ii\qB\tau},\nn\\
\eea
}
with
\begin{widetext}
\bea
\mathcal{I}_1(\tau,m,0)&\equiv&\int\frac{d^2\ell_\parallel}{(2\pi)^2}\frac{\left[\ell_0(\ell_0+2m)-\ell_3^2+2m^2\right]\left[m-\left((\ell_0+m)\gamma^0+\ell_3\gamma^3\right)\Op{\downarrow}\right]e^{\ii\tau\ell_\parallel^2}}{\left[\ell_0(\ell_0+2m)-\ell_3^2\right]^2\sqrt{\ell_0(\ell_0+2m)}}\nn\\
&=&-\ii\int\frac{d^2\ell_\text{E}}{(2\pi)^2}\frac{\left[\ell_4(\ell_4-2\ii m)+\ell_3^2-2m^2\right]\left[m-\left((\ii\ell_4+m)\gamma^0+\ell_3\gamma^3\right)\Op{\downarrow}\right]e^{-\ii\tau\ell_\text{E}^2}}{\left[\ell_4(\ell_4-2\ii m)+\ell_3^2\right]^2\sqrt{\ell_4(2\ii m-\ell_4)}},
\eea
\end{widetext}
where in the second line we performed a Wick rotation in order to get the Euclidean space, i.e., $\ell_0\to\ii\ell_4$, and $\ell_\text{E}\equiv\ell_4^2+\ell_3^2$.

With the following change of variables

\bea
\ell_4&=&r\sin\theta\nn\\
\ell_3&=&r\cos\theta,
\label{eq:r_theta}
\eea
\bea
&&\mathcal{I}_1(\tau,m,0)=-\ii\int_0^\infty \frac{r\,dr}{2\pi}f_1(m,r)e^{-\ii\tau r^2},
\eea
where
{\small
\bea
f_1(m,r)&\equiv&\int_0^{2\pi}\frac{d\theta}{2\pi}\frac{2m^2-r^2+2\ii m r\sin\theta}{r^{5/2}\left(r-2\ii m\sin\theta\right)^2\sqrt{\sin\theta\left(2\ii m-r\sin\theta\right)}}\nn\\
&\times&\Big[(m+\ii r\sin\theta)\gamma^0\Op{\downarrow}+r\cos\theta\,\gamma^3\Op{\downarrow}-m\Big].
\eea
}

In order to obtain analytical results, we consider a situation where the fermion mass is a small energy scale, such that $m^2\ll\qB$. This condition is physically accessible, given that in heavy-ion collisions, the produced magnetic fields are estimated to be about the pion mass squared. Therefore, for light quarks, this is a good approximation. So, by expanding in a power series around $m=0$ we get:
\bea
f_1(m,r)&\approx&\int_0^{2\pi}\frac{d\theta}{2\pi}\Bigg\{\frac{\ii\left(\cos\theta\,\gamma^3\Op{\downarrow}+\ii\sin\theta\,\gamma^0\Op{\downarrow}\right)}{r^2\left|\sin\theta\right|}\nn\\
&-&\Bigg[(3+\cot^2\theta)\cot\theta\,\gamma^3\Op{\downarrow}\nn\\
&+&\ii\left(\csc^2\theta+2\gamma^0\Op{\downarrow}\right)\Bigg]\left|\sin\theta\right|\frac{m}{r^3}\Bigg\}+O(m^2)\nn\\
&=&\frac{2\ii\left(\ln(2)+2\gamma^0\Op{\downarrow}\right)m}{\pi r^3}+O(m^2).
\eea

Hence, 
\bea
&&\mathcal{I}_1(\tau,m,0)=\frac{\left(\ln(2)+2\gamma^0\Op{\downarrow}\right)m}{\pi^2}\int_0^\infty dr\frac{e^{-\ii\tau r^2}}{r^2}. \nn\\
\eea

It is clear that the radial integral has a singularity when $r\to0$. To avoid it, we regularize this by the following prescription:
\bea
\int_0^\infty dr\frac{e^{-\ii\tau r^2}}{r^2}\to\int_0^\infty dr\frac{e^{-\ii\tau r^2}}{r^2+\mu^2}
\eea
so that, as is shown in Appendix~\ref{Ap:Details on the calculation of the regularized phase-space integral},
{\small
\bea
&&\mathcal{J}(\tau,\mu)\equiv\int_0^\infty dr\frac{e^{-\ii\tau r^2}}{r^2+\mu^2}\nn\\
&=&\frac{\pi e^{\ii\mu^2\tau}}{2\mu}\left[1-(1+\ii)C\left(\sqrt{\frac{2\tau}{\pi}}\mu\right)-(1-\ii)S\left(\sqrt{\frac{2\tau}{\pi}}\mu\right)\right],\nn\\
\eea
}
where $S(z)$ and $C(z)$ are the Fresnel integrals. With this in mind, the singularity is isolated by power expanding around $\mu=0$, i.e.,
\bea
\mathcal{J}(\tau,\mu)=\frac{\pi}{2\mu}-(1+\ii)\sqrt{\frac{\pi\tau}{2}}+O(\mu),
\eea
then, we defined the regularized integral as:
\bea
\mathcal{J}^r(\tau,\mu)\equiv\mathcal{J}-\frac{\pi}{2\mu},
\eea
so that
\bea
\int_0^\infty dr\frac{e^{-\ii\tau r^2}}{r^2}\to\lim_{\mu\to0}\mathcal{J}_R(\tau,\mu)=-(1+\ii)\sqrt{\frac{\pi\tau}{2}}.\nn\\
\label{eq:int_regularizada}
\eea

Putting all together:
{\small
\bea
&&\lim_{\substack{p_0\to m\\\mathbf{p}\to 0}}\left[-\ii\Sigma_{\Delta}^{(1)}(p)\right]^r\nn\\
&=&-\frac{3\ii(1+\ii)\left(\ln(2)+2\gamma^0\Op{\downarrow}\right)\alpha_\text{em}\Delta\qB^2m}{\pi^3}\nn\\
&\times&\sqrt{\frac{\pi}{2}}\int_0^\infty d\tau\frac{\sqrt{\tau}e^{-\epsilon\tau}}{2+\ii\qB\tau},
\eea
}
where the superscript $r$ stands for {\it regularized}. 

Now, note that:
\bea
&&\sqrt{\frac{\pi}{2}}\int_0^\infty d\tau\frac{\sqrt{\tau}e^{-\epsilon\tau}}{2+\ii\qB\tau}\nn\\
&\approx&-\frac{\ii\pi}{\qB\sqrt{2\epsilon}}+\frac{(1+\ii)\pi^{3/2}}{\sqrt{2}\qB^{3/2}}+O(\epsilon^{1/2}),
\label{eq:integral_tau_regularizada}
\eea
in such a way that we also absorb the singularity of $\epsilon\to 0^+$ in the counterterms associated to the noise-averaged self-energy. Therefore:
\bea
&&\lim_{\substack{p_0\to m\\\mathbf{p}\to 0}}\left[-\ii\Sigma_{\Delta}^{(1)}(p)\right]^r\nn\\
&=&\frac{3\sqrt{2}}{\pi^{3/2}}\left(\ln(2)+2\gamma^0\Op{\downarrow}\right)\alpha_\text{em}\Delta\sqrt{\qB} m+O(m^2).\nn\\
\eea

Finally, from the completeness relation
\bea
\Op{\uparrow}+\Op{\downarrow}=\mathbb{1},
\eea
we can split the expression in the two spin-components as follows:
\bea
\lim_{\substack{p_0\to m\\\mathbf{p}\to 0}}\left[-\ii\Sigma_{\Delta}^{(1)}(p)\right]^r=-\ii\widetilde{\Sigma}_{\Delta}^{(1,\downarrow)}\Op{\downarrow}-\ii\widetilde{\Sigma}_{\Delta}^{(1,\uparrow)}\Op{\uparrow},\nn\\
\eea
where
\begin{subequations}
    \bea
    -\ii\widetilde{\Sigma}_{\Delta}^{(1,\downarrow)}\equiv\frac{3\sqrt{2}}{\pi^{3/2}}\left(\ln(2)+2\gamma^0\right)\alpha_\text{em}\Delta\sqrt{\qB} m,
    \eea
    and
    \bea
    -\ii\widetilde{\Sigma}_{\Delta}^{(1,\uparrow)}\equiv\frac{3\sqrt{2}}{\pi^{3/2}}\ln(2)\alpha_\text{em}\Delta\sqrt{\qB} m.
    \eea
\end{subequations}

The next structure is
 \bea
    -\ii\Sigma_{\Delta}^{(2)}(p)=-4\alpha_\text{em}\Delta\qB\int\frac{d^4k}{(2\pi)^4}\frac{\Theta_2(k)\gamma^3\Op{+}}{(k-p)^2+\ii\epsilon},\nn\\
    \eea
which after following the same procedure has the form:
{\small
\bea
-\ii\Sigma_{\Delta}^{(2)}(p)&=&12\ii\alpha_\text{em}\Delta\qB\int_0^\infty d\tau\exp\left[-\frac{2\ii\tau\pt^2}{2+\ii\tau\qB}-\epsilon\tau\right]\nn\\
&\times&\int\frac{d^2\ell_\perp}{(2\pi)^2}\exp\left[-\left(\frac{2+\ii\qB\tau}{\qB}\right)\ell_\perp^2\right]\nn\\
&\times&\int\frac{d^2\ell_\parallel}{(2\pi)^2}\frac{
(\ell_3+p_3)e^{\ii\tau\ell_\parallel^2}\gamma^3\Op{+}}{\left[\left(\ell_\parallel+p_\parallel\right)^2-m^2\right]\sqrt{\left(\ell_0+p_0\right)^2-m^2}}.\nn\\
\eea
}
so that in the limits $ p_0\to m, \mathbf{p}\to0$ it vanishes by the parity of the parallel integration, i.e.,
\bea
\lim_{\substack{p_0\to m\\\mathbf{p}\to 0}}\left[-\ii\Sigma_{\Delta}^{(2)}(p)\right]=0.
\eea

The third structure is given by:
{\small
\bea
    -\ii\Sigma_{\Delta}^{(3)}(p)=-4\alpha_\text{em}\Delta\qB\sign\int\frac{d^4k}{(2\pi)^4}\frac{\Theta_3(k)\slashed{k}_\parallel\ii\gamma^1\gamma^2}{(k-p)^2+\ii\epsilon},\nn\\
    \eea
    }
so that with the described procedure it turn to the expression: 
{\small
\bea
-\ii\Sigma_{\Delta}^{(3)}&=&4\ii\alpha_\text{em}\Delta\qB\sign\nn\\
&\times&\int_0^\infty d\tau\exp\left[-\frac{2\ii\tau\pt^2}{2+\ii\tau\qB}-\epsilon\tau\right]\nn\\
&\times&\int\frac{d^2\ell_\perp}{(2\pi)^2}\exp\left[-\left(\frac{2+\ii\qB\tau}{\qB}\right)\ell_\perp^2\right]\nn\\
&\times&\int\frac{d^2\ell_\parallel}{(2\pi)^2}\frac{\left(\slashed{l}_\parallel+\slashed{p}_\parallel\right)e^{\ii\tau\ell_\parallel^2}\ii\gamma^1\gamma^2}{\left[\left(\ell_\parallel+p_\parallel\right)^2-m^2\right]\sqrt{\left(\ell_0+p_0\right)^2-m^2}}.\nn\\
\eea
}

Given the parity of the integrand, the odd terms in $\ell$ doesn't contribute to the integral. Hence, we have
\bea
&&\lim_{\substack{p_0\to m\\\mathbf{p}\to 0}}\left[-\ii\Sigma_{\Delta}^{(3)}(p)\right]\nn\\
&=&\frac{\ii\alpha_\text{em}\Delta\qB^2\sign}{\pi}\int_0^\infty d\tau\frac{\mathcal{I}_3(\tau,m,0)e^{-\epsilon\tau}}{2+\ii\qB\tau},\nn\\
\eea
where
\bea
&&\mathcal{I}_3(\tau,p_0,p_3)\nn\\
&\equiv&\int\frac{d^2\ell_\parallel}{(2\pi)^2}\frac{(\ell_0+m)e^{\ii\tau\ell_\parallel^2}\gamma^0(\ii\gamma^1\gamma^2)}{\left[\left(\ell_\parallel+p_\parallel\right)^2-m^2\right]\sqrt{\left(\ell_0+p_0\right)^2-m^2}}.\nn\\
\eea

After a Wick rotation and performing the change of variables of Eq.~\eqref{eq:r_theta}:
{\small
\bea
\mathcal{I}_3(\tau,m,0)&=&-\ii\int\frac{d^2\ell_\text{E}}{(2\pi)^2}\frac{(\ii\ell_4+m)e^{-\ii\tau\ell_\text{E}^2}\gamma^0(\ii\gamma^1\gamma^2)}{\left[\ell_4(\ell_4-2\ii m)+\ell_3^2\right]\sqrt{\ell_4(2\ii m-\ell_4)}}\nn\\
&=&-\ii\int_0^\infty\frac{rdr}{2\pi}f_3(m,r)e^{-\ii\tau r^2}
\eea
}
with
{\small
\bea
f_3(m,r)\equiv\int_0^{2\pi}\frac{d\theta}{2\pi}\frac{(\ii r\sin\theta+m)\gamma^0(\ii\gamma^1\gamma^2)}{r^{3/2}\left(r-2\ii m\sin\theta\right)\sqrt{\sin\theta\left(2\ii m-r\sin\theta\right)}},\nn\\
\eea
}
so that around $m=0$:
\bea
f_3(m,r)&\approx&\int_0^{2\pi}\frac{d\theta}{2\pi}\Bigg[\frac{\sin\theta}{|\sin\theta|r^2}+\frac{2\ii|\sin\theta|m}{r^3}\Bigg]\gamma^0(\ii\gamma^1\gamma^2)\nn\\
&+&O(m^2)\nn\\
&=&\frac{8\ii m}{2\pi r^3}\gamma^0(\ii\gamma^1\gamma^2)+O(m^2).
\eea

Then,
\bea
\mathcal{I}_3(\tau,m,0)&=&\frac{2m}{\pi^2}\int_0^\infty dr\frac{e^{-\ii\tau r^2}}{r^2}\gamma^0(\ii\gamma^1\gamma^2)\nn\\
&\to&-\frac{2(1+\ii)m}{\pi^2}\sqrt{\frac{\pi\tau}{2}}\gamma^0(\ii\gamma^1\gamma^2),
\eea
where the prescription of Eq.~\eqref{eq:int_regularizada} was used, and after using Eq.~\eqref{eq:integral_tau_regularizada}, we get:
\bea
&&\lim_{\substack{p_0\to m\\\mathbf{p}\to 0}}\left[-\ii\Sigma_{\Delta}^{(3)}(p)\right]^r\nn\\
&=&\frac{2\sqrt{2}}{\pi^{3/2}}\alpha_\text{em}\Delta\sign \sqrt{\qB} m\gamma^0(\ii\gamma^1\gamma^2)+O(m^2).\nn\\
\eea

Finally, from the identity
\bea
\ii~\sign\gamma^1\gamma^2=\Op{\downarrow}-\Op{\uparrow}
\eea
the expression can be written as
\bea
\lim_{\substack{p_0\to m\\\mathbf{p}\to 0}}\left[-\ii\Sigma_{\Delta}^{(3)}(p)\right]^r=-\ii\widetilde{\Sigma}_{\Delta}^{(3)}\left(\Op{\downarrow}-\Op{\uparrow}\right),\nn\\
\eea
where
\bea
-\ii\widetilde{\Sigma}_{\Delta}^{(3)}\equiv\frac{2\sqrt{2}}{\pi^{3/2}}\alpha_\text{em}\Delta\sqrt{\qB} m\gamma^0.
\eea

From the latter, the final expression for $\ii\Sigma_{\Delta}(p)$ in the limits $p_0\to m$, and $\mathbf{p}\to0$ is:
\bea
\lim_{\substack{p_0\to m\\\mathbf{p}\to 0}}\left[-\ii\Sigma_{\Delta}(p)\right]^r=-\ii\widetilde{\Sigma}_{\Delta}^{(\downarrow)}\Op{\downarrow}-\ii\widetilde{\Sigma}_{\Delta}^{(\uparrow)}\Op{\uparrow},
\eea
with
\begin{subequations}
{\small
    \bea
    -\ii\widetilde{\Sigma}_{\Delta}^{(\downarrow)}\equiv\frac{\sqrt{2}}{\pi^{3/2}}\left[3\ln(2)+8\gamma^0\right]\alpha_\text{em}\Delta\sqrt{\qB} m,\nn\\
    \eea
}
and
\bea
    -\ii\widetilde{\Sigma}_{\Delta}^{(\uparrow)}\equiv\frac{\sqrt{2}}{\pi^{3/2}}\left(3\ln(2)-2\gamma^0\right)\alpha_\text{em}\Delta\sqrt{\qB} m.\nn\\
    \eea
\end{subequations}

Finally, we can define two projections given by
\bea
\mathcal{P}^{(\pm)}\equiv\frac{1}{2}\left(\mathbb{1}\pm\gamma^0\right),
\eea
so that
\bea
\mathcal{P}^{(+)}+\mathcal{P}^{(-)}&=&\mathbb{1},\nn\\
\mathcal{P}^{(+)}-\mathcal{P}^{(-)}&=&\gamma^0,
\eea
in such a way that we can split the self-energy contribution into four subspaces, namely
\bea
\lim_{\substack{p_0\to m\\\mathbf{p}\to 0}}\left[-\ii\Sigma_{\Delta}(p)\right]_r=\sum_{\sigma=\uparrow,\downarrow}\sum_{\lambda=\pm1}\left[-\ii\widetilde{\Sigma}_{\Delta}^{(\sigma,\lambda)}\Op{\sigma}\mathcal{P}^{(\lambda)}\right],\nn\\
\eea
where we defined the coefficients:
\begin{subequations}
   \bea
-\ii\widetilde{\Sigma}_{\Delta}^{(\downarrow,\pm)}\equiv\frac{\sqrt{2}}{\pi^{3/2}}\left(3\ln(2)\pm8\right)\alpha_\text{em}\Delta\sqrt{\qB} m,
\eea 
   \bea
-\ii\widetilde{\Sigma}_{\Delta}^{(\uparrow,\pm)}\equiv\frac{\sqrt{2}}{\pi^{3/2}}\left(3\ln(2)\mp2\right)\alpha_\text{em}\Delta\sqrt{\qB} m.
\eea 
\end{subequations}

\section{Details on the calculation of the regularized phase-space integral}\label{Ap:Details on the calculation of the regularized phase-space integral}

In previous sections along this appendix, we arrived at the regularized integral
\bea
\mathcal{J}(\tau,\mu) = \int_0^{\infty} dr \frac{e^{-\ii \tau r^2}}{r^2 + \mu^2 }
\label{eq:Itaumu}
\eea

Let us first apply the following integral tranformation of the denominator in Eq.~\eqref{eq:Itaumu}
\bea
\frac{1}{r^2 + \mu^2} = \int_0^{\infty}dy\, e^{-y(r^2 + \mu^2)}.
\eea

Substituting into Eq.~\eqref{eq:Itaumu}, and integrating over $r$ first, we obtain
\bea
\mathcal{J}(\tau,\mu)&=& \int_0^{\infty}dy\, e^{-\mu^2 y} \int_0^{\infty} dr\, e^{-(y + \ii\tau) r^2}\nn\\
&=& \frac{\sqrt{\pi}}{2}
\int_0^{\infty} dy\, e^{-\mu^2 y} (y + \ii \tau)^{-1/2}
\eea

Let us shift the integration variable, by defining $z = y + \ii\tau$, to arrive at
\bea
\mathcal{J}(\tau,\mu)&=& \frac{\sqrt{\pi}}{2} \int_{\ii \tau}^{\infty} dz\, e^{-\mu^2 (z - \ii\tau)}z^{-1/2}\nn\\
&=& e^{i\mu^2\tau}\frac{\sqrt{\pi}}{2\mu}\int_{i\mu^2\tau}^{\infty}dz\,e^{-z}\,z^{-1/2},
\eea
where in the second step we re-scaled the integration variable $z\rightarrow z/\mu^2$.
Let us now define the auxiliary variable
\bea
z = v^2\Longrightarrow z^{-1/2}dz = v^{-1}\cdot 2v dv =  2 dv
\eea
and hence the integral becomes
\bea
\mathcal{J}(\tau,\mu) &=& e^{i\mu^2\tau} \frac{\sqrt{\pi}}{\mu} \int_{\mu\sqrt{\ii \tau}}^{\infty} dv\,e^{-v^{2}}\nn\\
&=& e^{i\mu^2\tau} \frac{\sqrt{\pi}}{\mu}\left( \int_0^{\infty}dv\,e^{-v^{2}}
- \int_{0}^{\mu\sqrt{\ii\tau}}dv\,e^{-v^2}\right)\nn\\
&=& e^{i\mu^2\tau} \frac{\pi}{2\mu}\left(
1 - \frac{2}{\sqrt{\pi}}\int_0^{\mu\sqrt{\ii\tau}}dv\,e^{-v^2}
\right)\nn\\
&=& e^{i\mu^2\tau} \frac{\pi}{2\mu}\left(
1 - \Phi\left( \mu\sqrt{\ii\tau}  \right)\right),
\label{eq:Itaumu2}
\eea
where in the last line we applied the integral representation of the probability integral $\Phi(z)$.
In particular, when $z = \sqrt{\ii}x$ for $x\in\mathbb{R}$, the probability integral is related to the Fresnel integrals $C(z)$ and $S(z)$ by the identity
\bea
\Phi(\sqrt{\ii}x) = \sqrt{2\ii}\left( C(\sqrt{2/\pi}x) -\ii S(\sqrt{2/\pi}x) \right).
\eea

Applying this last identity, and using $\sqrt{\ii} = (1 + \ii)/\sqrt{2}$, we have
\bea
\Phi(\mu\sqrt{\ii\tau}) =  (1+\ii)C\left(\mu\sqrt{\frac{2\tau}{\pi}}\right)
+ (1-\ii)S\left(\mu\sqrt{\frac{2\tau}{\pi}}\right).
\eea

Substituting this last relation into Eq.~\eqref{eq:Itaumu2}, we finally obtain
\bea
\mathcal{J}(\tau,\mu) &=& e^{i\mu^2\tau} \frac{\pi}{2\mu}\Big[ 1 -  (1+\ii)C\left(\mu\sqrt{\frac{2\tau}{\pi}}\right)\nn\\
&-& (1-\ii)S\left(\mu\sqrt{\frac{2\tau}{\pi}}\right) \Big]
\eea

In the vicinity of $\mu = 0$, the integral reduces to the power series
\bea
\mathcal{J}(\tau,\mu) &=& \frac{\pi}{2\mu} - \left(1 + \ii \right)\sqrt{\frac{\pi\tau}{2}} + \frac{\ii \pi \tau\mu}{2} + O(\mu^2)\nn\\
\eea

The regularized integral is thus obtained by subtracting the pole at $\mu \rightarrow 0$, such that
\bea
\mathcal{J}^{r}(\tau,\mu) \equiv \mathcal{J}(\tau,\mu) - \frac{\pi}{2\mu}
\eea

Therefore, in the selfenergy terms, we can finally evaluate this regularized expression at $\mu \rightarrow 0$, such that
\bea
\int_0^{\infty} dr \frac{e^{-\ii \tau r^2}}{r^2 } \rightarrow 
\lim_{\mu\rightarrow 0} \mathcal{J}^{r}(\tau,\mu)
= - \sqrt{\pi \tau}\frac{\left(1 + \ii \right)}{\sqrt{2}}.\nn\\
\eea

\end{document}